\documentclass[twocolumn,superscriptaddress]{revtex4-1}

\usepackage{color}
\usepackage{graphicx}
\usepackage{amssymb}
\usepackage{amsmath}
\usepackage{epstopdf}
\usepackage{natbib}
\usepackage{tabularx}

\usepackage[T1]{fontenc}
\usepackage{times}

\usepackage[text={7.25in,10in},centering]{geometry}

\usepackage{afterpage}
\usepackage{layouts}

\graphicspath{
{Figs/}
}

\begin{document}

\title{Genetic correlations greatly increase mutational robustness and can both reduce and enhance evolvability}

\author{Sam F. Greenbury}
\affiliation{Theory of Condensed Matter Group, Cavendish Laboratory, University of Cambridge, UK}
\author{Steffen Schaper}
\affiliation{Rudolf Peierls Centre for Theoretical Physics, University of Oxford, UK}
\author{Sebastian E. Ahnert}
\affiliation{Theory of Condensed Matter Group, Cavendish Laboratory, University of Cambridge, UK}
\author{Ard A. Louis}
\affiliation{Rudolf Peierls Centre for Theoretical Physics, University of Oxford, UK}

\begin{abstract}
\noindent
Mutational neighbourhoods in genotype-phenotype (GP) maps are widely believed  to be more likely to share characteristics than expected from random chance.   Such \textit{genetic correlations} should
strongly influence evolutionary dynamics. 
We explore and quantify these intuitions  by comparing three GP maps -- a model for RNA secondary structure, the HP model for  protein tertiary structure, and the Polyomino model for protein quaternary structure -- to a simple random null model that maintains the number of genotypes mapping to each phenotype, but assigns genotypes randomly.  The mutational neighbourhood of a genotype in these GP maps is much more likely to contain genotypes mapping to the same phenotype than in the random null model.  Such  {\em neutral correlations} can be quantified by the robustness to mutations, which can be many orders of magnitude larger than  that of the null model, and crucially, above the critical threshold for the formation of  large neutral networks of mutationally connected genotypes which enhance the capacity for the exploration of phenotypic novelty.  Thus neutral correlations increase evolvability.
We also study {\em non-neutral correlations}: Compared to the null model,  i)   If a particular (non-neutral) phenotype is found once in the 1-mutation neighbourhood of a genotype, then the  chance of finding that phenotype multiple times in this neighbourhood is larger than expected; 
  ii) If two genotypes are connected by a single neutral mutation, then their  respective non-neutral 1-mutation neighbourhoods are more likely to be similar; iii) If a genotype maps to a folding or self-assembling phenotype, then its non-neutral neighbours are less likely to be a potentially deleterious non-folding or non-assembling phenotype.   Non-neutral correlations of type i) and ii) reduce the rate at which new phenotypes can be found by neutral exploration, and so may diminish evolvability, while non-neutral correlations of type iii) may instead facilitate evolutionary exploration and so increase evolvability.
\end{abstract}
\maketitle


\medskip

\section{Author Summary}
Evolutionary dynamics arise from the interplay of mutations acting on genotypes and natural selection acting on phenotypes. Understanding the structure of the genotype-phenotype (GP) map is therefore critical for understanding evolutionary processes.  We address a simple question about structure:  Are the genotypes positively correlated?  That is, will the mutational neighbours of a genotype be more likely to map to similar phenotypes than expected from random chance?  John Maynard Smith and others have argued that the intuitive  answer is yes.  
 Here we quantify these intuitions by comparing model GP maps for RNA secondary structure, protein tertiary structure, and protein quaternary structure to a random GP map.  We find strong neutral correlations: Point mutations are orders of magnitude more likely than expected by random chance to link genotypes that map to the same phenotype, which  vitally increases the potential for evolutionary innovation by generating neutral networks. If GP maps were uncorrelated like the random map, evolution may not even be possible.
   We also find correlations for non-neutral mutations: Mutational neighbourhoods are less diverse than expected by random chance. Such local heterogeneity slows down the rate at which new phenotypic variation can be found.  But non-neutral correlations also enhance evolvability by lowering the probability of mutating to a deleterious non-folding or non-assembling phenotype.

\section{Introduction}
In a classic paper~\cite{smith1970natural},  published in 1970, John Maynard Smith introduced several key ideas for describing the structure of genotype-phenotype (GP) maps.  He first outlined the concept of a protein space, the set of all possible sets of amino acid chains, and suggested that for evolution to smoothly proceed, these should be connected as networks of functional protein phenotypes that can be interconverted by (point) mutations.  He then argued that one criterion for such networks to exist is for a protein $X$ to have at least one mutationally accessible neighbour which is ``meaningful, in the sense of being as good or better than $X$ in some environment''.  In other words, if $X$ has $N$ mutational neighbours, then the frequency $f$ of ``meaningful'' proteins in its mutational neighborhood should satisfy $f > 1/N$.  He pointed out that this was likely to be true in part due to the ubiquity of neutral mutations, which had  been famously  proposed by  Kimura~\cite{kimura1968evolutionary} and King and Jukes~\cite{king1969non}  just a few years prior to his paper.  But he also gave a second reason for expecting connected networks, namely that, ``There is almost certainly a higher probability that a sequence will be meaningful if it is a neighbour of an existing functional protein than if it is selected at random.'' This concept that mutational neighbours  differ from the random expectation is what we will call {\em genetic correlations}.

Following Maynard Smith, many authors have explored the role of networks of genotypes connected by single point mutations.  Lipman and Wilbur~\cite{lipman1991modelling} first showed that large networks of mutationally connected genotypes mapping to the same phenotype are found in the  Hydrophobic-Polar (HP) model for protein folding, introduced by Dill~\cite{dill1985theory,lau1989lattice}.  They also pointed out that neutral mutations allow a population to traverse these networks, facilitating access to a larger variety of alternate phenotypes. Schuster and colleagues~\cite{schuster1994sequences} developed these themes further using detailed models for the secondary structure of RNA~\cite{hofacker1994fast}.  They coined the term ``neutral network''  to describe sets of mutationally connected genotypes that map to the same secondary structure phenotype. As RNA secondary structure is fairly easy to calculate and thermodynamics based models such as the Vienna package are thought to provide an accurate prediction of real RNA secondary structure \cite{mathews2004incorporating,reuter2010rnastructure}, the nature of neutral networks in these models has been extensively studied~\cite{schuster1994sequences, hofacker1994fast,fontana2002modelling,cowperthwaite2008ascent,aguirre2011topological,schaper2014arrival,wagner2005robustness,wagner2011origins}.  Since these pioneering works, neutral networks have been considered in GP maps of other biological processes, including models for gene networks~\cite{ciliberti2007robustness,raman2011evolvabilitysig}  metabolic networks~\cite{samal2010genotype}  and the Polyomino model for self-assembling protein quaternary structure~\cite{greenbury2014tractable}.

\begin{figure*}[t]
\centering
	\includegraphics{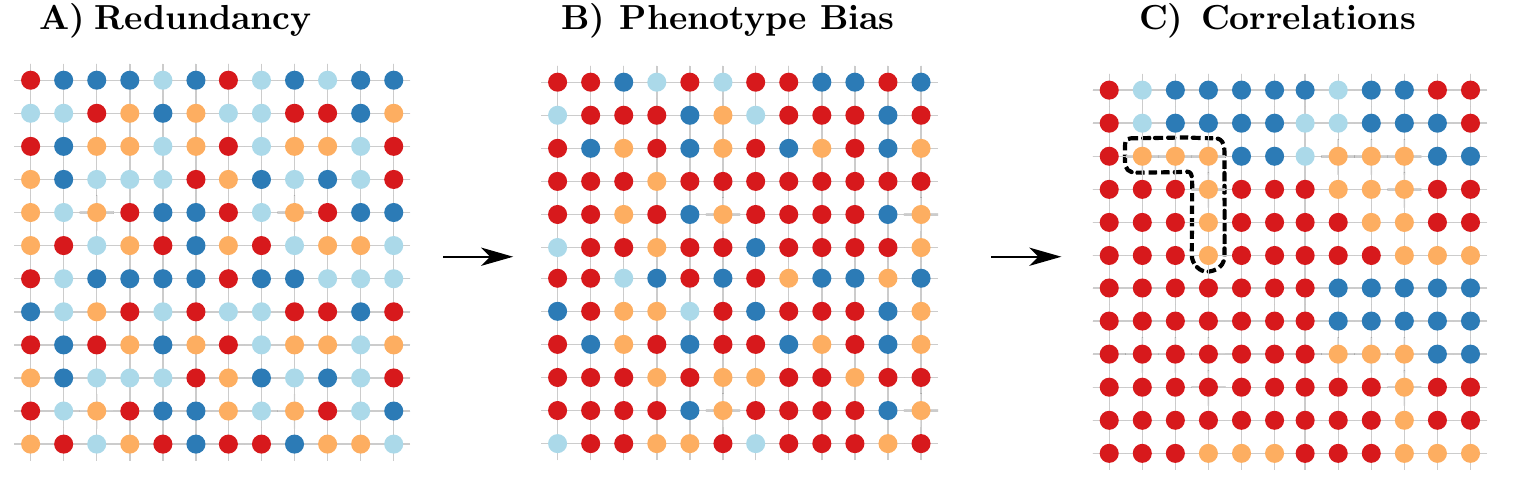}
	\caption{\textbf{Schematic depiction of the  GP map properties of redundancy, phenotype bias and neutral correlations.} Phenotypes are represented by colours, genotypes as nodes and mutations as edges. A) Each colour appears  multiple times with uniform redundancy. B) Some colours appear more often than others, demonstrating a phenotype bias. C) A rearrangement of the colours from the middle plot illustrates positive neutral correlations where the same colours are more likely to appear near each other than would be expected by random chance arrangement. The black box surrounding the six orange genotypes depicts a single component (a set of genotypes connected by neutral point mutations, also called a neutral network) of the orange phenotype.  Such positive neutral correlations enhance the probability that  such neutral networks occur.
}
	\label{fig1}
\end{figure*}

 From these studies of model systems a number of basic principles have emerged, much of which has been reviewed in important books by Wagner~\cite{wagner2005robustness,wagner2011origins}. Firstly, for neutral networks to exist, the GP map should exhibit {\em redundancy}, where multiple genotypes  map onto the same phenotype. This many-to-one nature of the mappings is illustrated in  Fig.~\ref{fig1}A. Redundancy is of course closely linked to the existence of neutral mutations~\cite{kimura1968evolutionary,king1969non}, although the relationship between these concepts is not entirely unambiguous.   In the theory of neutral evolution, a mutation may lead to a slightly different phenotype, but as long as the change in fitness is small enough not to be visible to selection, it is considered to be effectively neutral~\cite{ohta1973slightly}. Whether selection can act depends on the degree of phenotypic change, the environment, and other factors such as the population size and mutation rate. Therefore, identifying whether or not a mutation is neutral can be complex, and the answer may vary as parameters external to the GP map change with time.    So while redundancy only couples identical phenotypes, and so is a more restrictive concept than neutral mutations, it  has the advantage of sidestepping  the subtle issues listed above and is therefore more easily applicable to the study of a static GP map.

The second basic principle to emerge is that  the number of genotypes per  phenotype (the redundancy) can vary, leading to {\em phenotype bias}, as depicted in Fig.~\ref{fig1} B. 
Thirdly,  it is generally the case that the larger the redundancy, the greater the mean mutational robustness  of genotypes mapping to that phenotype. Fourthly, the larger the neutral network, the greater the variety of alternative phenotypes within one (non-neutral) point mutation of the whole neutral network, leading to a positive correlation between robustness and measures of evolvability that count the number of different phenotypes that are potentially accessible~\cite{wagner2008robustness}.

 Finally, a key principle emphasised by Maynard Smith~\cite{smith1970natural}, but which  has earlier roots in concepts such as the shifting balance theory of Sewall Wright~\cite{wright1932fitnesslandscapes}, is that neutral mutations allow a population to access, over time,  a wider variety of potential alternative phenotypes than would be available around a single genotype~\cite{lipman1991modelling,fontana2002modelling,wagner2011origins}.   Evidence for the key role of these networks in promoting evolutionary innovation has been found, for example, in experiments on RNA structures~\cite{schultes2000one,hayden2011cryptic} and transcription factors~\cite{payne2014robustness}.

The main focus of this paper is genetic correlations. To explore and quantify how they affect concepts such as neutral networks, robustness and evolvability,  we study genetic correlations in three of the  GP maps mentioned above. These are  the sequence to RNA secondary structure map and  HP model for protein folding (tertiary structure), which have been extensively studied, as well as the more recently introduced Polyomino model for self-assembling protein quaternary structure.  Several  properties of these three GP maps have recently been compared~\cite{ferrada2012comparison,greenbury2014tractable} and we summarise some of the similarities and differences between them in the methods section. However, a detailed investigation of their genetic correlations has not yet been considered.

A key question we consider is how to define genetic correlations in a quantitative way.  For that one needs some kind of uncorrelated null model for how genotypes are distributed over phenotypes with which to compare the full biophysical systems.  We employ a model we call the {\em random GP map}. It has the same number of genotypes mapping to each phenotype as the biophysical GP map to which it is being compared, as well as the same basic type of genotype space (alphabet size and genome length), and nodes (genomes) that are linked by single mutations if they differ by one locus. The difference is that the genotypes are randomly distributed across the genotype space.  Of course one does not expect biological systems to have such a random distribution, but it is not at all straightforward to think of a better null expectation. The great advantage of having such a null model, even if one knows that it is has limitations,  is that it allows us to quantitatively contrast how the biophysical genotypes are organised across the genetic space, which should shed light on nature of the correlations that Maynard Smith introduced.

The paper is organised as follows. We first define our models in the methods section. Next we examine {\em neutral correlations}, schematically illustrated in  Fig.~\ref{fig1} C, through considering various measures of robustness that quantify the relative likelihood that mutationally neighbouring genotypes possess the same phenotype.  We then perform a similar analysis comparing the biological GP maps to the random map for {\em non-neutral correlations}.  Since these different kinds of correlations all modulate the way that novel variation arises through random mutations, we finish by commenting on how correlations affect subtle interplay of robustness and evolvability~\cite{wagner2005robustness,wagner2011molecular},  and also briefly suggest a few other forms of correlation that could be studied in GP maps.

\section{Methods}
\subsection{Biological GP maps:  RNA, Polyomino and HP models}
We consider three separate GP maps for low-level self-assembling biological systems.
Firstly, a model for RNA secondary structure \cite{hofacker1994fast} that determines which bases in an RNA sequence form bonded pairs. Secondly, the HP lattice model for protein tertiary structure \cite{dill1985theory, lau1989lattice}, that determines the three-dimensional shape of a folded amino acid chain.  And thirdly, the Polyomino model for protein quaternary structures\cite{ahnert2010self,johnston2011evolutionary,greenbury2014tractable}, where quaternary structure of proteins is the topological arrangement of separate folded amino acid chains.
All three of these models have been previously compared in ref.~\cite{greenbury2014tractable}, and  below we  briefly outline the three different systems.

\subsubsection{Vienna package for RNA secondary structure}
In the widely studied RNA secondary structure GP map~\cite{schuster1994sequences,hofacker1994fast,fontana2002modelling,cowperthwaite2008ascent,aguirre2011topological,schaper2014arrival,wagner2005robustness,wagner2011origins}, the genotypes are sequences made of an alphabet of four different nucleotides, and phenotypes are the secondary structures, which describe the bonding pattern in the folded structure with the lowest free energy for the given sequence.   
 Here, we use the  popular Vienna package~\cite{hofacker1994fast}, which uses an empirical free-energy model and dynamic programming techniques to efficiently find the lowest  free energy structures. We use Version 1.8.5 with all parameters set to their default values.    The RNA GP map  for length $L$ is referred to as RNA$L$. We present results from RNA12, with $4^{12} \approx 16.8 \times 10^{6}$ genotypes mapping to $57$ folded pheontypes,   RNA15 with $4^{15}=1.07 \times 10^9$ genotypes mapping to $431$ folded phenotypes and   and RNA20, with $4^{20} \approx 1.10 \times 10^9$ genotypes mapping to $11,218$ folded secondary structure phenotypes.
 
\subsubsection{HP lattice model for protein tertiary structure}
The HP lattice model represents proteins as a linear chain of either hydrophobic (H) or polar (P) amino acids  on a lattice~\cite{dill1985theory}. A simple interaction energy function is used between non-adjacent molecules in the chain. Hydrophobic-hydrophobic interactions are energetically favourable. We use an implementation with the interaction energy $E$ between the different potential pairs being classified as $E_\text{HH}=-1$, with $E_\text{HP} = E_\text{PP}=0$, as widely chosen by other authors, see e.g.~\cite{irback2002enumerating,ferrada2012comparison}.  A phenotype is defined for each shape (fold) on the lattice that is the unique free energy minimum of at least one sequence.  If a sequence has more than one structure as its minimum, then it is considered not to fold properly, and so is categorised as  belonging to the (potentially deleterious) general non-folding phenotype. 
The compact  HP model restricts the possible folds to those which are maximally compact, in an attempt to capture the globular nature of \textit{in vivo} proteins~\cite{li1996emergence}. We make use of both compact and non-compact HP GP maps by considering both the GP map for all folds of length, $L=24$ (denoted HP24) and all compact folds on the $5\times5$ square grid (of length $L=25$, denoted HP5x5).  For the non-compact HP24 GP map there are $2^{24} \approx 16.8 \times 10^{6}$ genotypes and  $61,086$ folded phenotypes, while for the compact HP5x5 model the $2^{25} \approx 33.6 \times 10^6$ genotypes map to a much smaller set of $549$  unique folded phenotypes.

\subsubsection{Polyomino model for protein quaternary structure}
Protein quaternary structure describes the topological arrangement of separate folded proteins that self-assemble into a well-defined cluster.   The Polyomino GP map is a recently introduced  lattice model  which, in the spirit of the HP lattice model for tertiary structure, provides a simplified but tractable GP map for  protein quaternary structure, as described in more detail in \cite{greenbury2014tractable}.   Very briefly,  it employs a set of $N_t$ square tiles with $N_c$ interface types, together with a set of rules that denote how the interfaces bind.  These sets are specified by genotypes in the form of linear strings. In this work, $N_c=8$ and we specify that the interface types interact in ordered odd-even pairs, such that the following interface types interact ($1\leftrightarrow2$, $3\leftrightarrow4$, $5\leftrightarrow6$).

The conversion of the tile set (genotype) into a bounded shape (phenotype) is achieved through simulating a lattice based self-assembly process. This begins with seeding where the first tile encoded in the genotype is placed on the lattice. Thereafter other tiles are randomly placed. If they form a bond they are kept, otherwise they are discarded. This assembly process is repeated until either no more bonds can be formed or else the number of tiles grows without limit.  For each set of tiles,  assembly is attempted several times. If a given  set of tiles self-assembles into a unique bounded shape, the genotype is considered to map to that shape (phenotype).   If on the other hand, the tile set does not always assemble to the same shape, or if it assembles to an unbounded shape (as occurs in sickle cell anaemia for example), then the genotype maps to the \textit{undefined phenotype} (UND).
  
The GP map resulting from $N_t$ kit tiles and $N_c$ interface types is denoted as $S_{N_t,N_c}$. In this work, we consider $S_{2,8}$  which has $1.7 \times 10^7$ genotypes mapping to $13$ different self-assembling phenotypes and the larger space $S_{3,8}$ which has $6.9 \times 10^{10}$ genotypes mapping to $147$ phenotypes. All parameters used for simulations were the same as in ref.~ \cite{greenbury2014tractable}.

\subsubsection{A deleterious phenotype in all three GP maps}
We also distinguish a \textit{deleterious phenotype} (del) in all three GP maps.  For the RNA  GP map, this occurs when the unbonded strand is the free-energy minimum, so that the strand does not fold. For the HP model this occurs when a sequence does not have a unique ground state structure, which is interpreted as the protein not folding.  For the Polyomino GP map this occurs when the set of tiles produces an UND phenotype. Note that, depending on the environment, many  other phenotypes may also be deleterious,  but the del phenotype will always be an evolutionary dead end.
 For RNA this del phenotype makes up $85\%$ of RNA12, $65\%$ of RNA15 and $33\%$ of RNA20. In the HP model the fraction is typically larger, consisting of $98\%$ of the HP24 map and $82\%$  for the HP5x5 mapping, while for the Polyomino GP map we find $54\%$ of $S_{2,8}$ and $80\%$ of $S_{3,8}$. For the RNA this fraction decreases with increasing genotype length $L$ (asymptotically the fraction $f_{del}$, of genotypes  mapping to the deleterious phenotype scales as $f_{del} \approx  21.4 \times 0.82^L$~\cite{dingle2015structure}) while for the Polyomino GP map $f_{del}$ increases for larger system size, and in the HP model the trend remains ambiguous~\cite{schram2013exact}.

\subsubsection{Similarities and differences in the three GP maps.}
Recently, direct comparisons between the properties of these GP maps have begun to be considered~\cite{ferrada2012comparison,greenbury2014tractable}. A common feature of all three GP maps is that they correspond to self-assembly processes of molecular structures in biological systems. Thermodynamics drives the assembly process in each case.   They differ in that  the RNA and HP model have linked units that are fixed in size, while the polyomino model has disconnected units that can assemble into multiple sizes.
 The GP maps share properties such as redundancy, bias and phenotypic robustness~\cite{ferrada2012comparison,greenbury2014tractable} (see Fig. \ref{fig1}), which are seen in a much wider range of GP maps~\cite{wagner2005robustness,wagner2011origins}.  However, the nature of these properties varies between the GP maps. For example,  the polyomino and RNA GP maps typically have fewer phenotypes and larger neutral networks than phenotypes in the HP model. 
Another key variable that differs between the GP maps is the base $K$. For RNA $K=4$, HP $K=2$ and for Polyominoes $K=8$.  These different genotypic topologies  as well as the way in basic units interact to generate phenotypes can affect the formation of neutral networks.   
For example, in RNA, a CG bond in a stem motif cannot be turned into a GC bond without breaking the bond~\cite{reidys1997generic}, a form of neutral reciprocal sign epistasis~\cite{schaper2011epistasis}.
For a secondary structure stem  motif of $n$ bonds, this phenomenon breaks the neutral set up into around $2^n$ separate components, often of similar size, that are connected by point mutations internally, but which need at least two mutations to be connected together.  These different components of the full neutral set are separate neutral networks. With $K=8$ and asymmetric bonding of bases ($1\leftrightarrow2$ and $2\leftrightarrow1$ being distinct), the Polyomino GP map may be similarly susceptible to this form of epistasis. The HP lattice model does not have this feature due to bonds being formed from non-adjacent neighbouring HH interactions.  How these connectivity patterns scale with increasing $L$ remains an open question, in part because the spaces grow exponentially with length, and so rapidly become much more difficult to comprehensively analyse.

\subsection{The random GP map as an uncorrelated null model}
As discussed in the Introduction, in order to quantify genetic correlations we must first define an uncorrelated null model  to which the biophysical GP maps can be compared. Here we employ a random GP map that was recently explicitly introduced for analysing whole GP map properties in ref.~\cite{schaper2014arrival}, but has also  been used implicitly in many earlier works~see e.g.~\cite{reidys1997generic,wagner2011origins,schuster2011problems}, although  we believe this is the first time this random model has been used to {\it define} correlations.

The random GP map shares the following properties with  the biological GP map to which it is being compared:  the same alphabet size $K$, genome length $L$, number of 1-mutation neighbours $(K-1)L$, number of genotypes $N_G = K^L$,  number of phenotypes $N_P$, and  frequencies $f_p$, defined as the fraction of all genotypes that possess phenotype $p$. We summarise the GP map nomenclature used in this paper in Table \ref{table1},  which compares what is shared and what is different between the biological and the random GP maps.

With these key global GP map properties fixed, the only difference between a biological GP map and its associated random GP map is that  the $F_p = f_p \times N_G $ genotypes for each phenotype $p$ are each randomly assigned to the set of $N_G$ possible genotypes. As phenotypes are randomly assigned, departures in properties between the two versions of the GP map may be considered to be due to {\em correlations}, that is , in contrast to the random GP map, the mutational neighbourhood of a genotype in the biological GP maps is affected, for example, by what phenotype it maps to.  We note that these correlations can be very complex, and depend not only on the identity of the phenotype, but also on higher order features such as the identity of one or two or more phenotypes in the direct neighbourhood (higher order correlations). It is almost certainly also true that, depending on the phenotype, the correlations may also depend on which of the $L$ genomic positions is mutated (see e.g. Maynard Smith's word game described in the discussion).   However, in this paper we mainly focus on the simplest kinds of correlations; for example for neutral correlations we mainly look at effects that are captured by the concept of robustness.

Other null models are conceivable. For example, in ref.~\cite{aguirre2011topological} the authors used an approach based on network theory~\cite{aguirre2011topological}, comparing the topology of neutral components to Erd\H{o}s-R\'enyi networks and scale-free networks. While this type of network is helpful for understanding the topology of neutral networks themselves, a focus of this work here is not just how genotypes of the same phenotype connect to each other, but how phenotypes are arranged in relation to each other.  Besides their simplicity, an advantage of using the random GP maps for this purpose is that the overall connectivity of the genotype space is left intact, along with several global properties of the map, allowing the way phenotypes are arranged to be directly considered.
Moreover, this random map has been used (implicitly and explicitly) throughout the literature, see e.g.~\cite{wagner2011origins,reidys1997generic,schuster2011problems}, and so it is of general interest to carefully analyse some of its properties.

 \begin{table}[t]
	\centering
	\begin{tabularx}{\linewidth}{X  r}
	\hline
	Properties shared by random and biological GP maps & Symbol \\
	\hline
		\hline
	 \textit{Alphabet size}: & $K$ \\
	\textit{Genotype length}: & $L$ \\
		Number of 1-mutation neighbours of a genotype: & $(K-1)L$ \\
	 Number of genotypes: & $N_G = K^L$ \\
	Number of  phenotypes: & $N_P$ \\
	 \textit{Redundancy}:  the size of neutral set or the number of genotypes that map to phenotype $p$ & $F_p$ \\
	\textit{Phenotype frequency}: the fraction of genotypes that map to  phenotype $p$,  & $f_p=\frac{F_p}{N_G}$ \\
	\hline
	\hline
	Properties that differ between random and biological GP maps & Symbol \\
	\hline
	\hline
	{\em Neutral set}: all genotypes that map to phenotype $p$ & $\mathcal{G}_p$ \\
	{\em Neutral component}: A subset of $\mathcal{G}_p$ that is fully connected by point mutations.  Also called a {\em neutral network}. & $\textrm{NN}$ \\
	The number of 1-mutation neighbours of genotype $g$ mapping to phenotype $p$  & $n_{p,g}$ \\
	\textit{Phenotype robustness}: mean robustness of all  genotypes mapping to a phenotype $p$ & $\rho_p$  \\
	 \textit{Phenotype mutation probability}: Probability that a point mutation from a genotype mapping to phenotype $p$ will generate a genotype mapping to phenotype $q$
	  &  $\phi_{qp} $ \\
	\hline
	\end{tabularx}
	\caption{\textbf{GP map nomenclature.} }
	\label{table1}
\end{table}

\section{Results}
\subsection{Phenotypic robustness and neutral correlations}
\begin{figure*}[t]
	\centering
	\includegraphics{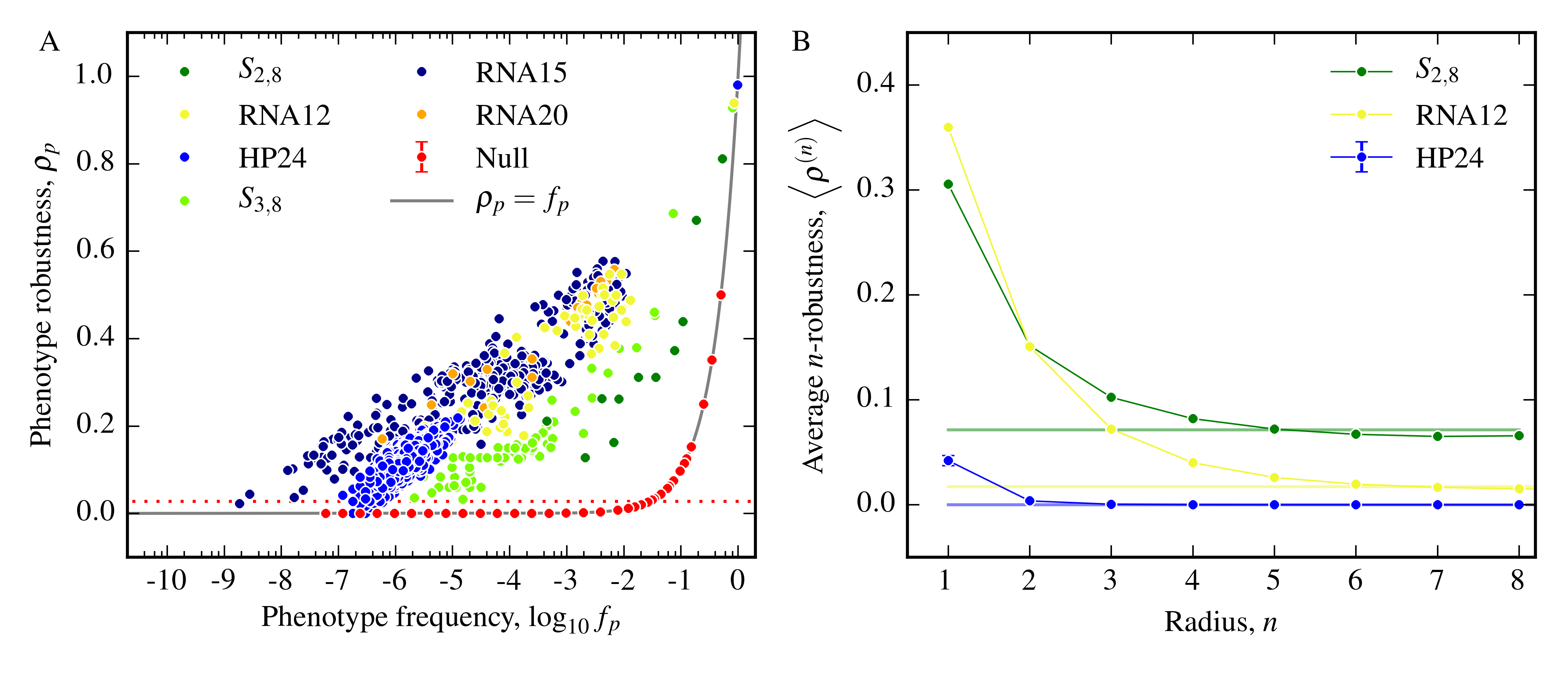} 
		\caption{\textbf{Greater mutational robustness indicates the presence of neutral correlations.} A) The phenotype robustness $\rho_p$ is plotted as a function of frequency $f_p$ for all phenotypes in the  RNA secondary structure models: RNA12, RNA 15, RNA20, the Polyomino models for protein quaternary structure: $S_{2,8}$ $S_{3,8}$  and the HP protein folding model HP24.  Each model has an associated random model with the same frequencies, but we only show one example, with $K=4$ and $L=12$ and a set of phenotypes chosen with a broad range of frequencies to best illustrate the relationship (red points). All random models closely follow the expected theoretical curve  $\rho_p = f_p$ (grey line).  The biophysical models exhibit a much larger robustness than the random models, which indicates the presence of positive neutral correlations. The red dotted line is $\delta$ (Eq. \eqref{eq:giant_component}) for $K=4$, $L=12$.  If ($\rho>\delta$) then large neutral networks are expected, which is much more likely for the biophysical models than for the random model.
B) The average $n$-robustness $\left< \rho^{(n)}\right>$, defined in Eq.~\protect\ref{eq:averagerobustness},  for each of the three biological GP maps, along with the expected values $\left< \rho^{(n)}\right> = 1/N_P$ for the associated random null models (flat coloured horizontal lines) is plotted against $n$. Across all three GP maps, we see a typical decay in robustness towards the random null model expectation with increasing mutational distance.  From this decay a neutral correlation length can be defined which is shorter for the HP model than for the other two models. Error bars for HP24 are the standard error on the mean of the average $n$-robustness.
}\label{fig2}
\end{figure*}
The concept of robustness to mutations is well established in the literature~\cite{wagner2005robustness,wagner2011origins}.  It is intimately tied to neutral correlations, in fact  robustness helps quantify the amount of neutral correlation present.  Before we study the more novel topic of non-neutral correlations, it is therefore interesting to compare various measures of robustness between the biological GP maps and the random uncorrelated GP map.

The 1-robustness of a single genotype $g$ that maps to phenotype $p$ is straightforwardly defined as the number of  genotypes $n_{p,g}$ that map to $p$ that are accessible within one point mutation of $g$. The  \textit{phenotypic robustness} $\rho_p$ of a phenotype $p$  is defined the average of the 1-robustness over the entire neutral set  $\mathcal{G}_p$~\cite{wagner2008robustness}. This can be expressed algebraically as
\begin{equation}\label{eq:1robustness}
\rho_p = \frac{1}{F_p} \sum_{g \in \mathcal{G}_p} \frac{n_{p,g}}{(K-1)L}
\end{equation}

In a random GP map, phenotypes are arranged randomly over genotypes so the probability that a genotype leads to phenotype $p$ is given by its frequency $f_p$, independently of  the identity of its neighbours.   The phenotypic robustness therefore is simply
$$\rho_p = f_p$$
and the mean number of  neutral neighbours is 
$$\left<n_{g,p}\right> = (K-1)Lf_p$$
which is the expectation value for a binomial distribution with $(K-1)L$ trials and probability of a given neighbour being $f_p$.  It is independent of the identity of the genotype $g$.

We define {\em neutral correlations} as the difference in how genotypes mapping to the same phenotype are distributed in a biologically relevant GP map as compared to the associated random GP map null model.  One way of characterising these neutral correlations is by comparing the phenotype robustness $\rho_p$ to the random expectation $\rho_p = f_p$. The violation of this equality is a sufficient (though not necessary) condition for the existence of neutral correlations.  
 Moreover, we define a phenotype $p$ to have  {\it positive} neutral correlations if $\rho_p > f_p$ is satisfied. This is intuitive -- when robustness is greater than $f_p$ then phenotypes are closer to each other in the genotype network than would be expected by random chance. 

Using the above definitions around neutral correlations, we explicitly consider the robustness in the various GP maps. In Fig.~\ref{fig2}A, we compare the phenotypic robustness across our three biological GP maps to the robustness of the associated random GP map. The figure confirms both the analytical result derived above for the random model that $\rho_p=f_p$ (we only show one schematic random map  in the figure, but the others have the same behaviour). In sharp contrast, for the biological GP maps we find that, very roughly, $\rho_p \propto \log f_p$, so that the robustness is much larger than would be expected for the null model, in fact by several orders of magnitude for smaller $f_p$.  Since $\rho_p \gg f_p$, this  indicates the presence of extremely strong neutral correlations in these biological GP maps.   Of course the fact that $\rho_p > f_p$ is not a new finding, but it is instructive to show this trend  displayed explicitly for entire mappings in the three kinds of biological systems.

\subsection{Generalised robustness and neutral correlations}
We next extend phenotype robustness to $n$-mutations. 
\textit{Generalised robustness} or \textit{$n$-robustness} $\rho^{(n)}_p$, measures phenotypic robustness for a greater number of mutations. It is defined as the  robustness of a genotype with phenotype $p$ to $n$ independent mutations to its genotype, rather than just the single mutation discussed above. This can be expressed algebraically as
\begin{equation}\label{eq:nrobustness}
\rho^{(n)}_p = \frac{1}{F_p}\sum_{g\in\mathcal{G}_p} n^{(n)}_{p,g} \frac{1}{{L\choose n}(K-1)^n}
\end{equation}
where $n^{(n)}_{p,g}$ is the number of $n$-mutant neighbours of $g$ with phenotype $p$ and the normalisation on the right-hand of the sum is the total number of $n$-mutants. In the same way as for the phenotype robustness, the $n$-robustness is averaged across the neutral set  $\mathcal{G}_p$ of all genotypes that map to phenotype $p$.

A further quantity we define is the \textit{average n-robustness} $\left<\rho^{(n)}\right>$ which is the average of the $n$-robustness over all phenotypes in a given GP map:
\begin{equation}\label{eq:averagerobustness}
\left<\rho^{(n)}\right>  = \frac{1}{N_P} \sum_{j \in \mathcal{P}} \rho^{(n)}_j
\end{equation}
where $\mathcal{P}$ is the set of all $N_P$ phenotypes in the GP map.  In contrast to the two previous definitions that measure robustness for a single phenotype,  it is a general property of the whole GP map.   One could imagine generalising this further to a subset of the phenotypes, for example those whose frequencies $f_p$ are greater than the average $N_P/N_G$.

To establish the $n$-robustness and average $n$-robustness in the random GP map, the same logic can be applied as in the previous section. Since the probability of finding a phenotype is uniformly distributed over the genotype space, the $n$-robustness is given by
$$
\rho_p^{(n)} = f_p
$$
with the $n$-robustness the same for all $n$, leading to an average $n$-robustness:
\begin{align}\label{eq:av_nrobustness}
\left<\rho^{(n)}\right> =& \frac{1}{N_P}\sum_{j \in \mathcal{P}} f_j \nonumber\\
				    =& \frac{1}{N_P}
\end{align}
since the phenotype frequencies in a  GP map sum to unity.
The inequality $\left<\rho^{(1)}\right>  \neq 1/N_P$ can be used to define whether a biological GP map possesses neutral correlations as a whole.

We consider the average $n$-robustness against the radius $n$ for the three GP maps $S_{2,8}$, RNA12 and HP24. A sample of 100 genotypes for each phenotype in the respective systems is taken (apart from HP24 where a sample of 100 randomly chosen phenotypes is made due to the large number of phenotypes) and the $n$-robustness is measured and averaged over phenotypes.
In Fig.~\ref{fig2}B, we plot the average $n$-robustness at each radius along with the flat expectation lines from Eq.~\ref{eq:av_nrobustness} for the null models. In all three cases we observe a decay from greater than the null values for small radii to slightly less than the null expectation at larger radii. 
The reason  for this drop below the random expectation can be understood intuitively:  given that positive neutral correlations are present, the over-representation for small radii must be balanced at larger radii by under-representation in order for the number of genotypes to balance.

We also define a {\em neutral correlation length} $n^\ast$ which measures the mutational hamming distance over which neutral correlations extend.  We define $n^\ast$ for a phenotype to be equal to the smallest value of $n$ where $\rho^{(n)}_p < f_p$ and for the GP map when $\left<\rho^{(n)}_p\right> < 1/N_P$.  We find that $n^\ast = 7$ for the RNA12 model,   $n^\ast = 6$ for the Polyomino $S_{2,8}$ model and $n^\ast = 5$ for the HP24 model. 
The neutral correlation length is smaller for the HP model than the other two systems. As discussed in the methods section, and illustrated in Fig.\ref{fig2}A, the HP model typically has phenotypes with smaller frequencies/robustness than the other two systems suggesting neutral networks that do not expand to the same diameter which would reduce the expected neutral correlation length. All three models are of fairly small genome length $L$,  so one should be careful of reading too much into the numerical values of these correlation lengths. However, it may very well be that this ordering of models will persist for larger $L$.

\subsection{The presence of positive neutral correlations/higher phenotype robustness results in larger and fewer neutral components}
\begin{figure*}[!t]
	\centering
	\includegraphics{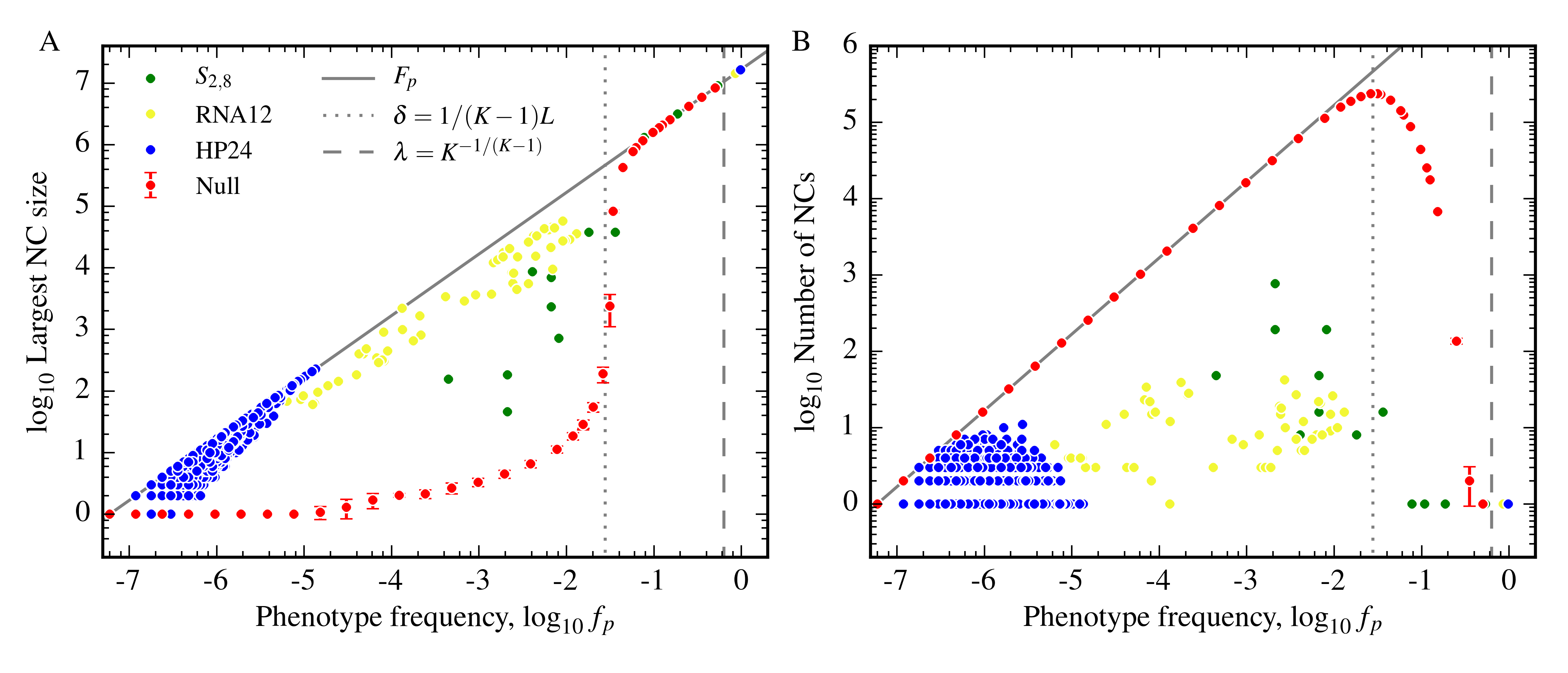}
	\caption{\textbf{Biological GP maps have much larger and fewer neutral components than their random counterparts due to neutral correlations.}
	A) The logarithm of the largest neutral component for a given phenotype  is plotted as a function of frequency  for random null models (with $K=4$, $L=12$) and three biological GP maps, RNA12, $S_{2,8}$ and HP24.  The vertical dotted line denotes the giant component threshold $\delta \approx 1/36$, defined in Eq.~\eqref{eq:giant_component}, for the schematic random model with $K=4,L=12$. The vertical dashed line denotes the single component threshold $\lambda \approx 0.37$, defined in Eq.~\eqref{eq:lambda}, for the schematic random model.  The biological GP maps show much larger connected components below these thresholds, due to the presence of positive neutral correlations.
	B) The logarithm of the total number of neutral components against frequency is plotted for the same models. The theoretical thresholds $\delta$ and $\lambda$ work well for random model but again the number of components in the biophysical models differ greatly from the random model expectation due to the presence of correlations.
	 In both plots, error bars represent a single standard deviation from the $100$ independent realisations of the  random null model used to derive the neutral component statistics.}
	\label{fig3}
\end{figure*}

Having illustrated the concept of positive neutral correlations -- measured by (generalised) robustness greater than that of the random null model  -- we next show how other properties of neutral networks  are affected by their presence.

 The neutral set $\mathcal{G}_p$ is the set of all genotypes mapping to phenotype $p$.
A component is the subset of the neutral set $\mathcal{G}_p$ that is connected by single point mutations. We use this term because it is commonly used in graph theory to denote a set that is connected.    Although the literature can be somewhat ambiguous, with the term neutral networks sometimes referring to the neutral set, and sometimes to a neutral component, we take a neutral network to be synonymous to a neutral component in this paper 
because if we have only point mutations then a population can only explore a neutral component  and may not be able access the whole  neutral set. 
 
There are several reasons why  a neutral set  may not be fully connected by neutral point mutations. If the  genotypes are too diffusely spread out over the full genotype space, then they may be disconnected.  But  in some cases basic biophysical constraints, such as the neutral reciprocal sign epistasis described in the Methods section, also lead to fragmentation. 
 
We begin by comparing the size of  neutral components in the random null model  to those found in our biological GP maps. In the random model, there are two important threshold values: firstly, the \textit{giant component onset}, when a phenotype's components change from being largely isolated to forming larger connected clusters, and secondly, the \textit{single component onset} where virtually all genotypes are taken up by a single giant connected component. 

As each genotype has many neighbours, a simple mean-field-like approach from  \textit{percolation theory} for random graphs~\cite{newman2010networks} should be fairly accurate. This suggests that 
the giant component onset begins when the average number of neighbours of a given genotype with the same phenotype is approximately unity, which was also  the criterion used by John Maynard Smith~\cite{smith1970natural}. For the null model, where phenotypes are assigned to genotypes completely randomly, this reduces to an explicit threshold frequency
\begin{equation}\label{eq:giant_component}
\delta = \frac{1}{(K-1)L}
\end{equation}
such that we expect the giant components for phenotypes with $f_p \gtrsim \delta$.
It can be shown analytically in the limit $L \rightarrow \infty$~\cite{reidys1997generic} that there is another transition at 
\begin{equation} \label{eq:lambda}
 \lambda = 1 -  \frac{1}{K^\frac{1}{K-1}}
\end{equation}
where, for $f_p \gtrsim \lambda$, all the components coalesce into one single giant component, so that the neutral set should be (nearly) fully  connected. 
While the giant component threshold $\delta$ scales as $1/L$, so that it decreases for larger maps,  the single component threshold $\lambda$ from Eq.~\ref{eq:lambda} is independent of  genome length $L$,  and only varies with alphabet size. For example,  $\lambda =0.5$ for $K=2$ and $\lambda \approx 0.37$ for $K=4$.  These are large frequencies that are unlikely to be reached for more than a single phenotype in any realistic GP maps.

In Fig.~\ref{fig3}, we plot how the largest component size (left) and number of components (right) varies with frequency in both a null model ($K=4$, $L=12$) and  three GP maps $S_{2,8}$, RNA12 and HP24.   
We first focus on the simple schematic null model. Data is calculated by averaging over 100 independent realisations of the random mapping of genotypes to phenotypes in a way that preserves the frequencies.
The largest component size, and the number of components formed by the phenotype, are then measured.  These values are shown in Fig.~\ref{fig3} for an array of frequencies in the schematic  null GP map.
  Below the giant component onset $\delta \approx 1/36$, most genotypes are completely isolated -- the total number of neutral components scales with $f_p$.
 Around the giant component threshold $\delta$,  this scaling changes markedly, and instead the  size of the largest neutral components scales linearly with $f_p$ and takes up the majority of the genotypes in the neutral set.  The number of components continues to decline until $f_p$ exceeds the single component connectivity threshold  $\lambda \approx 0.37$, at which point there is just one component and the neutral set is completely connected. 

We next consider the biological GP maps relative to the behaviour exhibited by the null model.
Firstly, all three GP maps have much larger maximum neutral set sizes than the random model. This is not surprising, as Fig.~\ref{fig2}A shows that, due to positive neutral correlations,   $\rho_p > \delta$ for most phenotypes in each system ($\rho = \delta$ for $K=4$, $L=12$ is shown as a dotted red line in the plot).  Once the probability of having a neutral neighbour is above the $\delta$ threshold, we expect large networks.
 For HP24 and RNA12, the largest neutral component size clearly grows linearly with frequency, and so scales linearly with the size of the neutral set. For the Polyomino space $S_{2,8}$ this scaling is less evident, but the components are still much larger than their random counterparts would be.

Secondly, for all three models, the number of components does not vary much with $f_p$, in contrast to the random model where this number scales, as expected, with the neutral set size if $f_p \lesssim \delta$.
Since these components typically have robustness above $\delta$ or even $\lambda$, the reason there are still multiple  components must be due to biophysical constraints which are not present in the random model, such as the neutral non-reciprocal sign epistasis discussed earlier for RNA.  These effects are to first order independent of $f_p$ which explains why the number of components does not correlate with $f_p$.
In each of these three models the largest ``phenotype'' of all is the deleterious non-folding or non-assembling one. Its frequency exceeds the threshold $\lambda$ and its neutral set is fully connected. 

Differences between the three biophysical GP maps observed in Fig.~\ref{fig2} can be fairly easily explained by some of the differences highlighted in the methods section. For example, the number of phenotypes per genotype is largest in the HP model, and smallest in the Polyomino model, which explains why they group at different frequencies.    While the number of components in Fig.~\ref{fig2} B) is generally much lower in the biological models than it is in the random model,  the HP model has significantly fewer components than the RNA and Polyomino models do. This may be due to the fact that the HP model does not exhibit effects such as neutral reciprocal sign epistasis, which fragments neutral sets in the other two systems.

We conclude that due to positive neutral correlations and the concomitant higher robustness,  the biophysical models considered here have large neutral networks even for frequencies $f_p$ that are several orders of magnitude lower than the random model large component threshold $\delta$. The abject failure of the random model to predict the robustness and the neutral network size highlights the importance of neutral correlations in these systems.

\subsection{Non-neutral phenotype mutation probability}
\begin{figure*}[!t]
	\centering
	\includegraphics{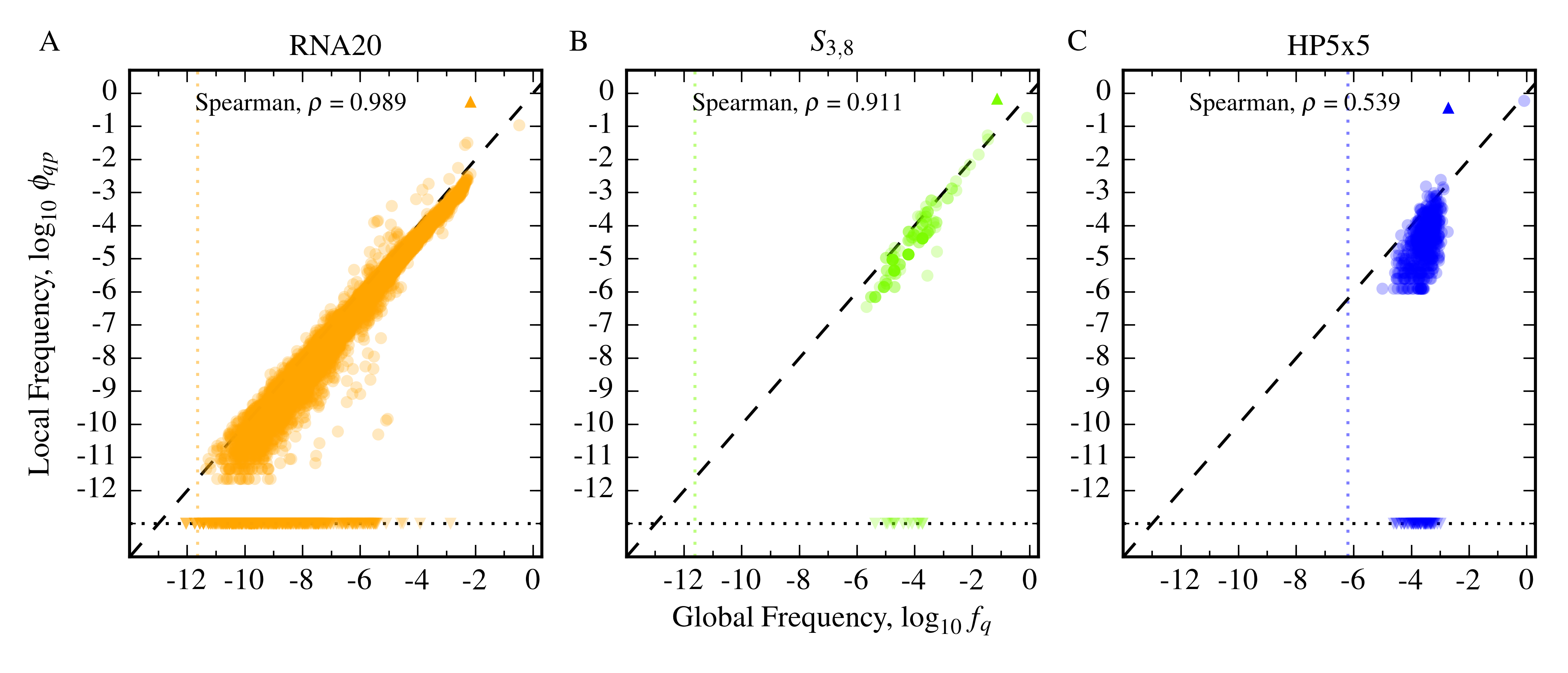}
	\caption{\textbf{Phenotype mutation probabilities scale with global frequency.} We present results for the three GP maps: A) RNA20, B) $S_{3,8}$ and C) HP5x5. We plot the relationship between $\phi_{qp}$ (circles) and $f_q$ for the largest non-deleterious phenotype $p$ in $S_{3,8}$ and HP5x5, and for the second largest in RNA20 (not the largest due to computational expense).  We see in each case a strong positive correlation ($p$-value $\ll 0.05$ in all cases), very similar to the expectation for the null model (not shown here, but for which the correlation is exact to within statistical fluctuations, see ref.~\cite{schaper2014arrival} and Appendix \ref{S2_Text}).  Spearman rank correlation coefficients are shown in the top-left of each plot.
 Differences from $\phi_{qp}=f_q$ are relatively small compared to the overall range of variation, except for sets of phenotypes that are not connected at all, which typically arise due to biophysical constraints. These are shown as downward triangles along the lower horizontal dotted line which represents $\phi_{qp}=0$.  For each plot, the upward triangle indicates $\phi_{pp}=\rho_p$, the phenotype robustness, which is always over-represented ($\rho_p \gg f_p$) due to neutral correlations.}
	\label{fig4}
\end{figure*}

We next consider non-neutral mutations.  The first question is:  Are two different phenotypes, on average, more or less likely to be connected to each other than one would expect by chance?
To address this question, we employ a generalisation of robustness, namely the  \textit{phenotype mutation probability} $\phi_{qp}$  of $q$ with respect to $p$, defined as  
the fraction of 1-point mutations of genotypes in the neutral set for phenotype $p$ that map to phenotype $q$.  This can be written as:
$$
\phi_{qp} =  \frac{1}{F_p(K-1)L}   \sum_{g \in \mathcal{G}_p} n_{q,g}. 
$$
Thus $\phi_{qp}$  averages a local property, $n_{q,g}$ -- the number of genotypes that map to phenotype $q$ found the 1-mutation neighbourhood of a genotype that maps to phenotype $p$ -- over the entire neutral set $\mathcal{G}_p$. Note that this phenotype mutation probability is not symmetric ($\phi_{qp} \ne \phi_{pq}$) and that, if $p=q$, it reduces to the phenotype robustness $\phi_{pp} = \rho_p $.
It has recently been shown~\cite{schaper2014arrival}  that $\phi_{qp}$ is a key quantity for incorporating the structure of a GP map into population genetic calculations.

In the null model we expect $\phi_{qp} = f_q$ to be an excellent approximation~\cite{schaper2014arrival}, with the caveat that it must be possible for enough genotypes to be sampled.  What do we mean by enough genotypes?  
Given a phenotype $p$ with redundancy $F_p$, there are at most $F_p (K-1)L$ unique neighbours available. This number provides an upper bound -- in reality, several neighbours of one genotype will also be neighbours of another genotype with the same phenotype, resulting in a reduction in the number of unique neighbours.  Nevertheless, this allows us to define a minimum threshold
\begin{equation}\label{eq:sample_threshold}
\gamma = \frac{1}{F_p (K-1) L}
\end{equation}
If $f_q \lesssim \gamma$, then the expected number of genotypes with phenotype $q$ found around phenotype $p$ is less than one, and the probability that $\phi_{qp} = 0$ due to statistical fluctuations becomes appreciable.   Further detail on how  $\phi_{qp}$ and $F_p$  relate when the threshold is not satisfied, which is mainly relevant  for smaller GP maps and for lower $F_p$, is provided in Appendix \ref{S1_Text}. Here we focus on phenotypes with larger $F_p$, in the larger GP maps of the previous section.

In Fig.~\ref{fig4}, we plot the relationship between the phenotype mutation probability $\phi_{qp}$ and global frequency $f_q$ around the RNA20 phenotype with the second largest neutral set, the assembling phenotype for $S_{3,8}$  with the largest neutral set, and the HP5x5 folding phenotype with the largest neutral set. For phenotypes in $S_{3,8}$ and HP5x5, with such large numbers of genotypes, all phenotypes have $f_q$ values that are significantly above $f_q=\gamma$ (vertical dotted lines), which is the approximate threshold at which at least one genotype of phenotype $q$ would be expected to be found. A small fraction of phenotypes lie close to the $f_q=\gamma$ threshold for RNA20, but by far the majority may be expected to be effectively sampled.
For RNA20  and $S_{3,8}$, we observe a very strong and highly significant positive correlation with the random null model expectation $\phi_{qp} = f_q$. 
In HP5x5, there is also a strong positive correlation, though less strong than in the RNA and Polyomino cases, with a greater number of phenotypes falling below the one-to-one expectation.  We did not plot the non-compact model HP24 because most of its frequencies are below the threshold $\gamma$ (see supporting information).

To summarise, in contrast to the robustness $\rho_p = \phi_{pp}$ where neutral mutations lead to strong deviations from the null model, the non-neutral phenotype mutation probabilities follow the random model expectation that $\phi_{qp} \approx f_q$ remarkably well. There are still important deviations, especially for those phenotypes that can not be reached due to biophysical constraints so that $\phi_{qp}=0$~\cite{schaper2011epistasis}. Moreover, it may be an interesting exercise to look more closely at phenotypes for which $\phi_{qp}$ is significantly greater or less than $f_q$ as such deviations could signal similarities or differences between phenotypes. For example, two RNA phenotypes with similar hairpin topology, but perhaps a difference of one bond in a stem may have a larger probability of interconverting than topologically more dissimilar RNA phenotypes. The difference between $\phi_{qp}$ and $f_q$ could then be used to quantify the difference between phenotypes $p$ and $q$. These more subtle types of correlation are beyond the scope of this paper.  At any rate, compared to the result in the previous sections showing the strength of neutral correlations, the dominant agreement with the random model is apparent. However, given that $\phi_{qp}$ is averaged over a neutral set, it may be that there are \textit{local} non-neutral correlations that are obscured by the averaging. With this in mind, we next investigate such local correlations.

\subsection{Non-neutral local over-representation correlations}

\begin{figure}[!t]
	\centering
	\includegraphics{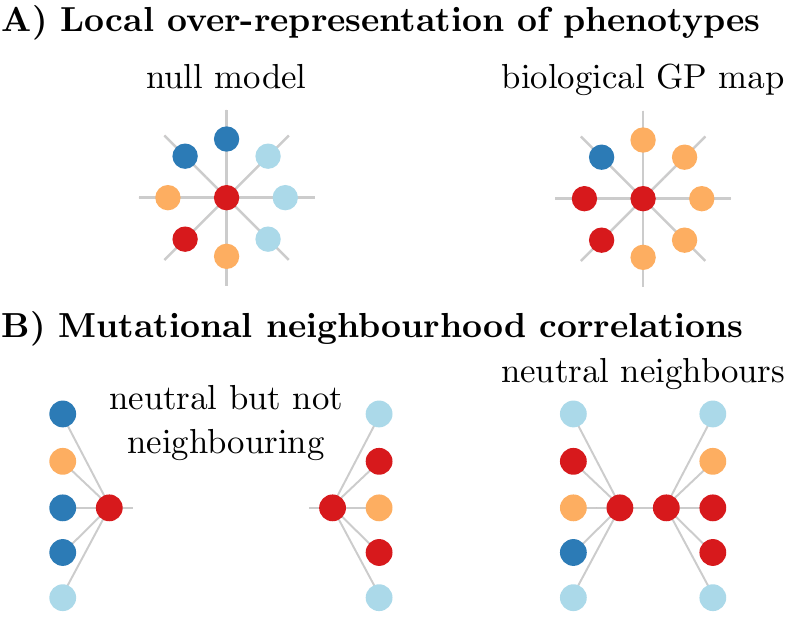}
	\caption{\textbf{Illustration of further non-neutral correlations.} A) On the right, the orange phenotype is over-represented relative to the null model: The red genotype in the centre has more orange neighbours than would be expected by the global frequency of orange.
	 B) The phenotypes that appear in the mutational neighbourhood of two neutral neighbours are expected to be more similar (right) than two non-neighbouring genotypes of the same phenotype (left).}
	\label{fig5}
\end{figure}

We first describe {\em non-neutral local over-representation correlations} which mean that, given phenotype $q$ is found in the 1-mutation neighbourhood of a genotype $g$ (which maps to phenotype $p \neq q$), then phenotype $q$ will appear a greater number of times in total than predicted by $f_q$ or $\phi_{qp}$ in this 1-mutation neighbourhood, as pointed out in ref.~\cite{wagner2008robustness}.  These correlations are illustrated in Fig.~\ref{fig5}A.  

 To measure 1-mutation neighbourhoods, we sample randomly chosen genotypes $g$ from  the neutral set $\mathcal{G}_p$, with a genotype of phenotype $q$ in its neighbourhood. We then measure the phenotype of all other neighbours of $g$. From this sample, we obtain the probability $P(q,p,m)$ of $q$ occurring $m$ times in the 1-mutation neighbourhood of a genotype mapping to phenotype $p$, given that $q$ occurs at least once.

Two control null expectations may also be derived for $P(q,p,m)$. In the random model where phenotypes are randomly assigned, given $q$ is in the 1-mutation neighbourhood of a genotype $g$ (at a specific genotype $g'$), the probability may be calculated as a binomial probability based upon the overall frequency of $q$, leading to
\begin{equation} \label{eq:null1}
P_1( q, p, m ) = {L(K-1) - 1 \choose m-1} f_q^{m-1}(1-f_q)^{L(K-1) - m}
\end{equation}

A second null expectation calculates the binomial probability  by replacing $f_q$ in Eq.~\ref{eq:null1} above by replacing the phenotype mutation probability $\phi_{qp}$ for the GP map instead:
\begin{equation}\label{eq:null2}
P_2( q, p, m ) = {L(K-1) - 1 \choose m-1} \phi_{qp}^{m-1}(1-\phi_{qp})^{L(K-1) - m}
\end{equation}
In contrast to $P_1(q,p,m)$, this form  accounts for any overall phenotypic heterogeneity known to be present in the GP map.

\begin{figure*}[!t]
	\centering
	\includegraphics{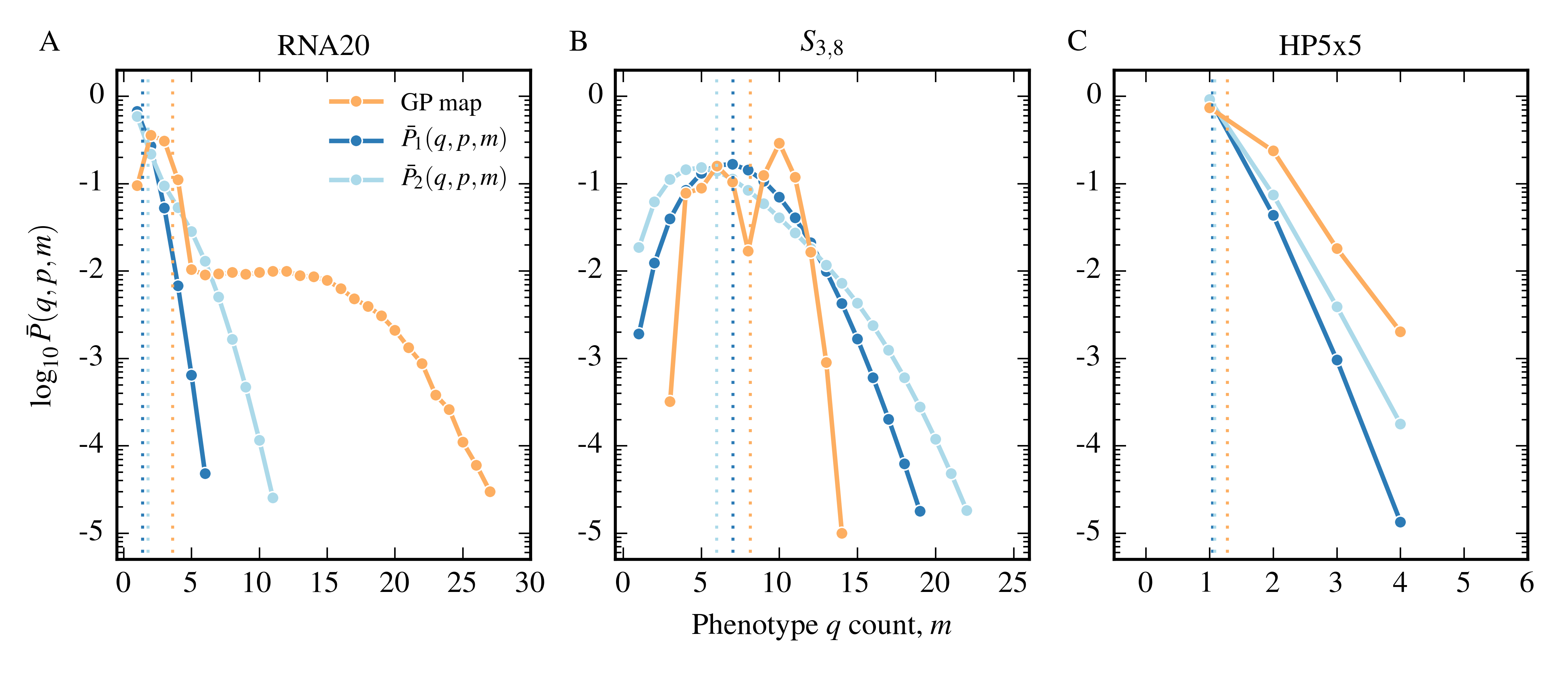}
	\caption{\textbf{Non-neutral local over-representation correlations result in phenotypes being more likely to be found multiple times around genotypes}. We present results for the three GP maps: A) RNA20, B) $S_{3,8}$ and C) HP5x5. We pick the same frequent phenotypes $q$  in each of our biological GP maps as used in Fig.~\ref{fig4}, and consider the prevalence of $q$ around genotype $g$ with phenotype $p$, \textit{given that $q$ occurs at least once} in the 1-mutation neighbourhood of $g$.  The average of $\bar{P}(q,p,m)$ across the $n=10$ most frequent phenotypes $p$ in the neighbourhood of $q$ (with $p\neq q$ and $p \neq \textrm{del}$), is compared to the respective averages for random null expectations $\bar{P}_1(q,p,m)$ and $\bar{P}_2(q,p,m)$ defined in the text. The mean of each distribution is plotted as a dotted line in each case.  Contiguous sections with a probability greater than $10^{-5}$ are joined with lines in order to guide the eye. The mean value of $m$  for each of the biological GP maps  and  the two random controls  are shown as respective dotted lines with the same colours.  Compared to the two null expectations of occurrence, $q$ is over-represented locally as demonstrated by the shift of the means  to the right.}
	\label{fig6}
\end{figure*}

We compare the actual local prevalence against these two null expectations in Fig.~\ref{fig6}. For RNA20, $S_{3,8}$ and HP5x5 we chose the same three phenotypes for  phenotype $q$ as we did in the previous section, while for phenotype $p$ we choose one-by-one the next $n=10$ largest (non-deleterious) phenotypes available in the GP map. By sampling 10,000  neighbourhoods for each of the $n=10$ phenotypes for $p$, we calculate an average for $P(q,p,m)$ across the phenotypes ($\bar{P}(q,p,m)$)
and compare this in Fig.~\ref{fig6} to the averages for the null expectations $\bar{P}_1(q,p,m)$ and $\bar{P}_2(q,p,m)$.
For each biological GP map, $q$ is more likely to  be over-represented, that is to appear multiple times if it appears at least once when compared to the null expectations, leading to a skewed distribution compared to the control case. The most striking result is seen in RNA20, where there is a substantial tail to the distribution.   We use average measures here to provide the general profile, smoothing out particular features that may occur between individual pairs of phenotypes $q$ and $p$, but the local over-representation is seen for any of the phenotype pairs considered.

  One consequence of these local over-representation correlations is that the probability that a genotype with phenotype $p$ has phenotype $q$ {\em at least once} in its 1-mutant neighbourhood is less than expected from $\phi_{qp}$. This is because those genotypes that have phenotype $p$ in their 1-mutant neighbourhood typically do so a greater number of times than expected. This must therefore be compensated with fewer genotypes around which phenotype $p$ actually appears at all, which we confirm numerically. In the RNA20 GP map, with the most frequent of the set of phenotypes used for $p$ and the next most frequent used as $q$, the probability of finding $q$ at least once is $0.12$ versus a null expectation of $1-(1-\phi_{qp})^{(K-1)L}=0.20$.  Thus these correlations lead to heterogeneity in the connections between phenotypes.

How these correlations affect evolutionary dynamics will depend on the  regime being explored~\cite{schaper2014arrival}.  If the population is neutrally exploring genotypes that map to phenotype $p$, then in the monomorphic regime of evolutionary dynamics, where $NL\mu \ll 1$, this heterogeneity will lead to a significant drop in the rate at which $q$ is  first discovered by neutral exploration. In the polymorphic regime where $N L \mu \gg 1$,  and different individuals  in the population have different genotypes, the rate at which novel variation with phenotype $q$ occurs may not be that different from the expectation given by $\phi_{qp}$, at least if the population is spread across a large enough number of different genotypes to average over local heterogeneity~\cite{schaper2014arrival}.

\subsection{Non-neutral local mutational neighbourhood correlations}
\begin{figure*}[!t]
	\centering
	\includegraphics{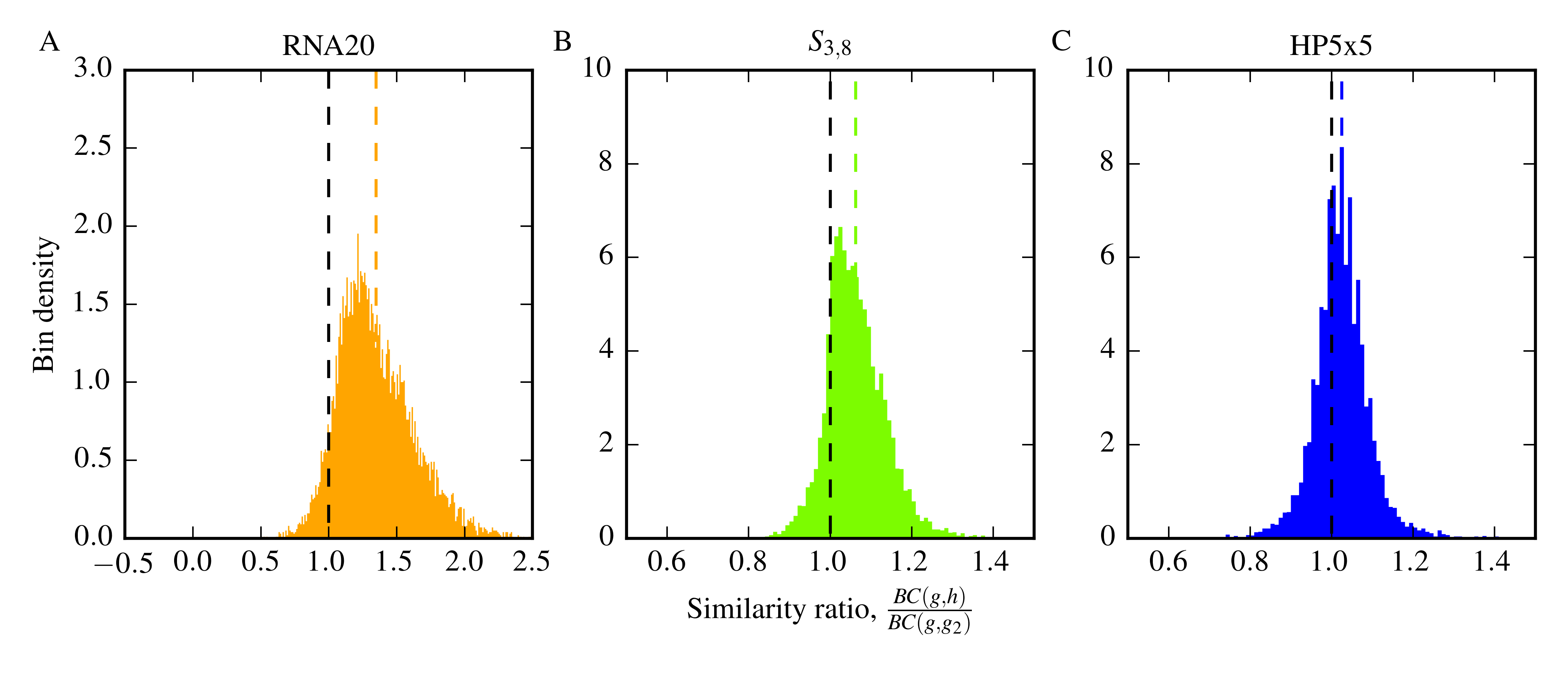}
	\caption{\textbf{Non-neutral local mutational neighbourhood correlations result in mutational neighbourhoods of neutral neighbours being more similar than randomly selected neutral pairs}. We present results for the three GP maps: A) RNA20, B) $S_{3,8}$ and C) HP5x5. Using the ratio of Bhattacharyya coefficients defined in Eq.~(\ref{eq:B}), we show that neutral neighbours ($g$ and $h$) have a closer phenotype probability distribution than a randomly chosen neutral pair ($g$ and $g_2$). This is seen through the ratio being skewed with a mean (coloured vertical dashed lines) larger than unity (black vertical dashed lines).   The standard error on this mean  is negligible compared to the distance of the mean from one. }
	\label{fig7}
\end{figure*}

\begin{figure*}[!t]
	\centering
	\includegraphics{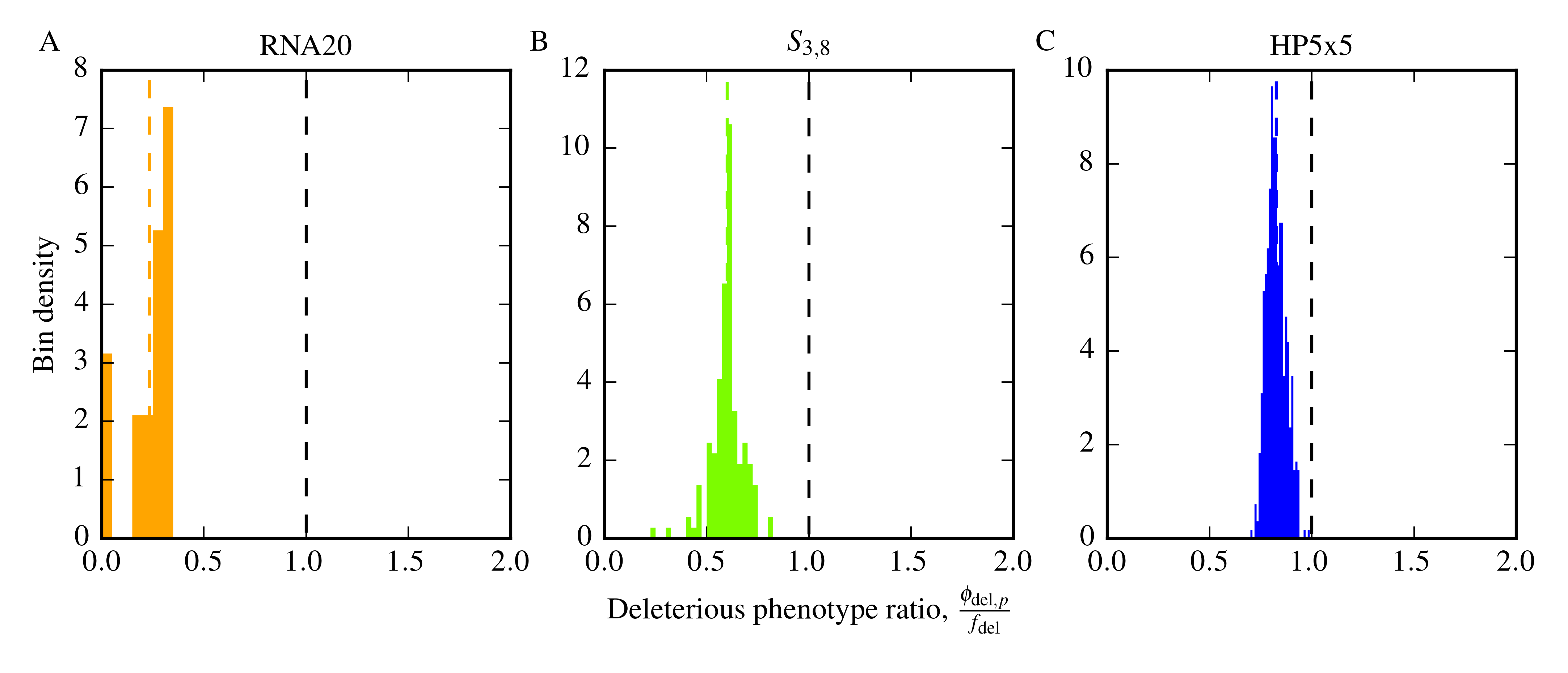}
	\caption{\textbf{Non-neutral deleterious phenotype correlations: The deleterious phenotype is under-represented in the neighbourhood of folding or self-assembling phenotypes.} We present results for the three GP maps: A) RNA20, B) $S_{3,8}$ and C) HP5x5. Histograms of the ratio  of the  phenotype mutation probability ($\phi_{\text{\textrm{del}},p}$) divided by the null model expectation of the global frequency ($f_\text{\textrm{del}}$) for the deleterious phenotype (non-folding for RNA/HP, non-assembling for Polyominoes). The distribution is clearly skewed to values $< 1$, as highlighted by the dashed vertical  coloured lines representing the mean in each case.
}
	\label{fig8}
\end{figure*}

We next examine \textit{non-neutral local mutational neighbourhood correlations}   which are illustrated schematically in Fig.~\ref{fig5}B. They show  that the 1-mutation neighbourhoods of two genotypes connected by a neutral point mutation are more likely to have similar phenotypic compositions than would be expected by two randomly chosen neutral non-neighbouring genotypes of the same phenotype. This type of correlation has already been demonstrated to exist for RNA~\cite{sumedha2007populations}.
To measure the similarity of neighbouring genotypes' mutational neighbourhoods, we  consider the  local quantity $\phi^\text{(local)}_{q,g}=n_{q,g}/(K-1)L$, which becomes $\phi_{qp}$ when averaged over the whole neutral set $\mathcal{G}_p$. We compare the $\phi^\text{(local)}_{q,g}$ for neighbouring genotypes with non-neighbouring genotypes in both the null model and biological GP maps.
  The similarity or difference could be measured in several different ways. The statistical  measure we employ here is the Bhattacharyya coefficient \cite{Bhattacharyya_1946}, which for two discrete probability distributions $x_i$ and $y_i$ may be expressed as
\begin{equation}\label{eq:B}
BC(x_i, y_i) = \sum_i \sqrt{x_i y_i}
\end{equation}
varying between 0 and 1 for maximally dissimilar and identical discrete probability distributions respectively.

To quantify whether neutral neighbours $g$ and $h$ have more similar phenotype distributions in comparison to non-neighbouring neutral genotype pairs $g$ and $g_2$, we compared the \textit{similarity ratio} of the Bhattacharyya coefficients, $BC(g,h)/BC(g,g_2)$, using the $\phi^\text{(local)}_{q,g}$ to define the distributions.  A ratio greater than unity indicates that the phenotype distributions around neutral neighbours are more similar than the randomly selected neutral pair, and vice versa.  We remove the $K-2$ mutual neighbours of $g$ and $h$ from the distributions as these will automatically contribute to similarity between the neighbourhoods in a trivial manner which we wish to exclude.

In Fig.~\ref{fig7} we plot histograms of the similarity ratio for 10,000 samples of $g$, $h$ and $g_2$ in RNA20, $S_{3,8}$ and HP5x5, where the phenotype sampled has the second largest frequency in RNA20, and the largest frequency in $S_{3,8}$ and HP5x5 (excluding the del phenotype).
 For 10,000 samples the means are $1.357\pm0.003$ for RNA20, $1.063\pm0.001$ for $S_{3,8}$ and $1.025\pm0.001$ for HP5x5, where the error is the standard error on the mean.  For
 RNA20 and $S_{3,8}$, a clear skew in the overall distribution may be visually observed, demonstrating that neutral neighbours, on average, have more similar mutational neighbourhoods. HP5x5 also has the mean of its distribution at a value slightly larger than unity but it is much more marginal in this case, and the skew is harder to detect. We note that in general, the non-neutral correlations are weakest for the HP5x5 model.    Finally,  just as is the case for the non-neutral local over-representation correlations of the previous section, these local mutational neighbourhood  correlations also reduce the rate at which novel phenotypes would be discovered by neutral exploration since a neutral neighbour is more likely to have some of the same phenotypes in its mutational neighbourhood, and so fewer alternatives.

\subsection{Non-neutral  deleterious phenotype correlations}
The final, and perhaps most important, type of non-neutral correlation we consider is the accessibility of the  deleterious phenotype from folding or self-assembling phenotypes, which we call \textit{non-neutral  deleterious phenotype correlations}.  This type of non-neutral correlation is closest to the type of correlation suggested by Maynard Smith~\cite{smith1970natural}.
 
 In Fig.~\ref{fig8} we plot histograms of the  ratio  $\phi_{\text{\textrm{del}},p}/f_\text{\textrm{del}}$ for all phenotypes $p$ in $S_{3,8}$ and HP5x5, and the top 20 most frequent (largest $f_p$) in RNA20 (limited due to computational expense of this larger system). In all cases, we see that the deleterious phenotype is significantly less frequent around the non-deleterious phenotypes. This behaviour contrasts to non-deleterious phenotypes, for which $\phi_{qp}  \approx f_q$. As a corollary of this effect, we also find $\rho_\textrm{del}/f_\textrm{del}$ equal to $1.10$, $1.16$ and $1.19$ for RNA12, $S_{3,8}$ and HP5x5 respectively, illustrating positive correlations, that is, a corresponding local over-representation of the deleterious phenotype in its own mutational neighbourhoods. Moreover, for $L=20$ we find that  $\rho_\textrm{del}/f_\textrm{del}=2.34$ suggesting that these positive neutral correlations may become stronger for larger $L$.

We find that the del phenotype forms only a single component in RNA12, $S_{2,8}$, HP24 and HP5x5.  This result is unsurprising because the large size of the del phenotype in each GP map (85\%, 54\%, 98\% and 82\% respectively) means that the frequencies are all well above the single component threshold $\lambda$ of  Eq.~\ref{eq:lambda}, which would lead to the expectation of a single component even in the random null model.

\section{Discussion}
In this paper, we have explored the role of genetic correlations, which we defined and quantified as the difference in how genotypes are mutationally connected for biologically relevant GP maps, compared to a random null model with the same global properties (alphabet size, genome length, and  number of genotypes per phenotype).  Genetic correlations provide a simple conceptual framework within which a number of topological properties of GP maps can naturally be captured.

\textbf{Neutral correlations:} We first explored the phenotype robustness for all three GP maps, showing that $\rho_p > f_p$ for all phenotypes,  a result which is not unexpected in the literature, but to our knowledge has not been compared for a set of whole GP maps before.  Since $\rho_p=f_p$ for the random model, the extent to which $\rho_p$ differs from $f_p$ can be viewed as a measure of  the extent of the neutral correlations.  

We also introduced the concept of $n$-robustness, which measures robustness over $n$ mutations.  From this we derived another criterion that measures the presence of neutral correlations by averaging this measure over all phenotypes and comparing to the null expectation: If   $\left< \rho^{(n)}\right>  > 1/N_p$ then there are positive neutral correlations.  We find that the enhanced probability of encountering a genotype mapping to the same phenotype can extend to multiple mutations $n$ away from genotypes. The extent of the correlations in sequence space can be quantified by the number of mutations $n^*$ at which the criterion is violated, a measure we call the correlation length of the neutral mutations.   We find that that $n^*$ is largest for the RNA12 model, and smallest for the HP24 model.   How $n^*$ or even the relative ordering of the correlation lengths between the different systems will scale with increasing genome length $L$ remains an open question.

It should also be emphasised that the full complexity of neutral correlations for a phenotype are only partly be captured with the measures we introduced here, which average over the entire neutral set.  As can be seen in~Fig 3A of ref.~\cite{schaper2014arrival}, a single neutral set can have  significant local heterogeneity in its internal connections.  Since neutral sets are frequently so vast that they cannot be fully explored by populations on evolutionary time-scales, such local heterogeneities may also have implications for evolutionary dynamics. Thus local measures of heterogeneity and robustness, which may be influenced, for example, by the identities of multiple neighbours, or the position of a mutation along a genome, may also be important to develop in future work. 

We found, for the three biological GP maps that we studied, that  the dominant relationship of robustness with frequency is $\rho_p \sim \log f_p$, a scaling that has already been pointed out earlier for RNA~\cite{jorg2008neutral,aguirre2011topological}. In an interesting paper that applies concepts from network theory~\cite{newman2010networks} to neutral sets, Aguirre et. al.~\cite{aguirre2011topological} rationalise this scaling for RNA by separating out the mutational behaviour of bound and unbound bases.  It would be interesting to see if a more general argument could be developed to explain the logarithmic scaling across all the systems we studied.  Moreover, these results also pose fascinating questions relating to why or how the constraints in a GP map lead to the kinds of neutral correlations they do.

\begin{figure*}[!t]
	\centering
	\includegraphics{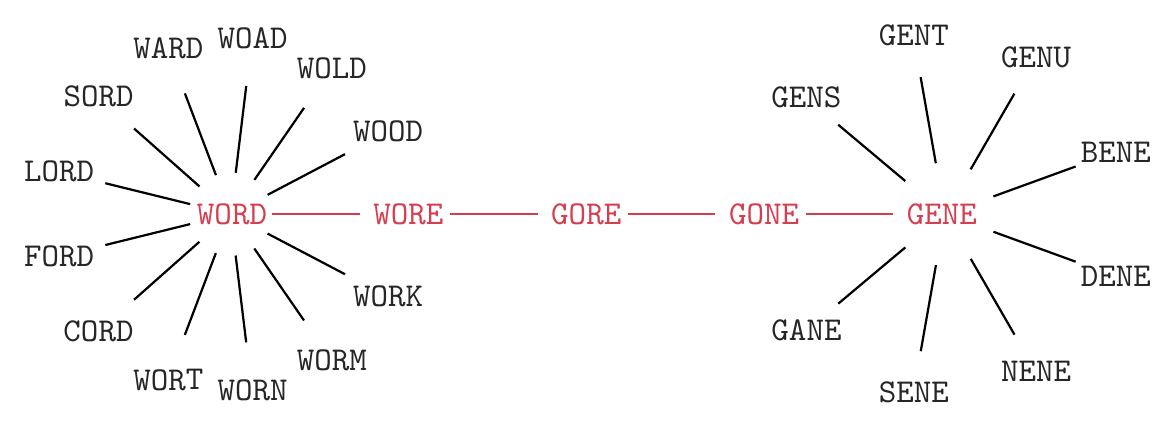}
	\caption{\textbf{Enhanced robustness in Maynard Smith's 4-letter word game.} 	
The single mutation path WORD $\rightarrow$ WORE   $\rightarrow$ GORE   $\rightarrow$  GONE  $\rightarrow$  GENE 
 is shown in red. All  valid words within a one letter mutation of  ``WORD'' and ``GENE'' are also depicted. According to the Merriam-Webster Official Scrabble Players Dictionary 2014, only  4,175 of the 456,976 possible 4-letter words are valid English words (at least for Scrabble).  Since each word has $100$ neighbours, for a random model, the expected number of valid words within a one letter mutation is $ < 1$.  Nevertheless, due to positive neutral  correlations, the probability that a {\it valid} word has another valid word as a 1-mutant neighbour is more than ten times greater, as illustrated above for ``WORD'' and ``GENE''. Such correlations make the game much easier. As pointed out by Maynard Smith, in a biological system, such correlations (in his case between ``meaningful sequences'') can facilitate evolutionary dynamics~\cite{smith1970natural}.
}
	\label{fig9}
\end{figure*}

Some clues to the underlying causes of neutral correlations and robustness can be gleaned from Maynard Smith's  original paper, where he illustrated the concept of a neutral network with the parlour game of  transitioning between connecting two words in the English language by changing one letter at a time, with each change also generating a valid word. He used the example of  changing ``WORD'' to ``GENE'' in four steps as illustrated in Fig. \ref{fig9}. There are $26^4 = 456,976$ different possible 4-letter words, but, according to the Merriam-Webster Official Scrabble Players Dictionary 2014, only $4,175$, or just over $0.9\%$, are valid English words. If we consider the set of valid words to be a phenotype, then it has frequency of only  $f_p=0.009$, just under the giant component threshold $\delta = 1/(K-1)L=0.01$, and well below the single component threshold $\lambda \approx 0.12$. On average the probability that a 4-letter word has a valid neighbour is just below one.  However, if we measure the phenotype robustness, that is the  mean probability that a valid 4-letter word has valid words in its $(K-1)L=100$ neighbours, we find that $\rho_p \approx 0.11$, or on average each word has 11 neighbours, which makes the game much simpler than if the random expectation held.

The reasons this game exhibits such a large enhancement of the robustness over  $f_p$ clearly arise from neutral correlations in English words.  For example, vowels are more likely to appear in specific places in words than would be expected by chance. The second letter of 4-letter English words has a 74\% chance of being a vowel compared to the 5/26 = 19\% overall average probability per locus. So if a word has a vowel placed at the second letter, it is much more likely to have neighbouring words using the same vowel, as can be seen in Fig.~\ref{fig9}.  This example illustrates how basic properties (in this case of language properties, in the case of our models, biophysical properties) can generate correlations, which can then result in  high robustness. 

A question often debated in the literature is the extent to which mutational robustness is selected for.
Here we argue that a major enhancement of robustness, often by many orders of magnitude over the random expectation, is not caused by selection, but  rather emerges from  the internal constraints of a GP map -- the way that genotypes map to phenotypes -- which naturally lead to positive neutral correlations.   It may still be the case that more robust genotypes can be selected for within a neutral set, or that these genotypes are favoured in certain dynamic regimes~\cite{van1999neutral}. It may also be true that in some cases a particular phenotype is preferred by selection because it is more robust than an alternative one.  But even if this is so, it is important to keep in mind that natural selection is still acting on variation that is already naturally quite robust due to correlations caused by biophysical constraints.
       
The relationship between robustness and evolvability has been the subject of much discussion in the literature~\cite{wagner2008robustness,cowperthwaite2008ascent,masel2009robustness,draghi2010mutational}.    Here we show,  as already anticipated by Maynard Smith~\cite{smith1970natural}, that if the phenotype robustness is roughly larger than $\delta =1/(K-1)L$, so that the expected number of neutral neighbours is greater than one, then the phenotype will exhibit large neutral networks.   In the random model, large networks will typically be very rare, but neutral correlations mean that robustness above the $\delta$ threshold is common for the biophysical GP maps. The effect can be very large. For example, for $L=55$ RNA, a recent study~\cite{dingle2015structure} suggests  that there are about  $N_P \approx 8 \times 10^{12}$ phenotypes, so that the mean frequency is  $\bar{f_p} \approx 10^{-13}$. In fact all phenotype frequencies are well below the threshold $\delta = 1/(3 \times 55) =0.00606$ above which we expect extended neutral networks.   On the other hand, the mean robustness of all phenotypes was estimated to be  $\bar{\rho_p} \approx 0.14 > \delta \gg \bar{f}_p$. Neutral correlations increase the probability of a nearest neighbour generating the same phenotype by on average about $12$ orders of magnitude over the mean expectation of the null model, lifting robustness well above the threshold $\delta$.     Thus the most important way that neutral correlations contribute to evolvability is by naturally creating robustness greater than the threshold needed to generate percolating networks which provide access to phenotypic novelty.   In fact it may very well be that without neutral correlations and its attendant robustness, evolution as we know it would not be possible
   
\textbf{Non-neutral correlations:} \textit{Non-neutral mutations} are important for the generation of novel variation.  For all three GP maps, the probability $\phi_{qp}$  that a phenotype $q$ is found by a point mutation from genotypes mapping to phenotype $p$ is, to first order, given simply by the global frequency: $\phi_{qp} \sim f_q$,  which is independent of $p$.

Since $f_q$ can span many orders of magnitude, the rate at which variation appears (which scales as $\tau_q \sim 1/\phi_{qp}~\sim 1/f_q$ if a population is neutrally exploring phenotype $p$~\cite{schaper2014arrival}) can also range over many orders of magnitude in these systems. These large differences can lead, both in the monomorphic and polymorphic regimes, to effects such as the {\em arrival of the frequent}~\cite{schaper2014arrival}, where  frequent phenotypes (with larger $f_q$)  fix in a population even when  alternate phenotypes that are much more fit, but much less frequent, are accessible in principle.

The reason these fitter phenotypes are not fixed is because they are unlikely to be found on evolutionary time-scales.  Natural selection can only work on variation that actually arises.  In the alternative case where  the system is effectively in steady state, so that a less frequent phenotype has a realistic probability to arise in a population, it can still be the case, especially at larger mutation rates, that a phenotype with lower fitness but larger frequency (and robustness) will fix, an effect known as the {\em survival of the flattest}~\cite{wilke2001digital}.

Finally, we note that $\phi_{qp}$ can be viewed as a non-neutral generalisation of the phenotypic robustness, but that $\phi_{pp} = \rho_p$ scales very differently with $f_p$ than $\phi_{qp}$ does when $p \neq q$. In the latter case local correlations more or less cancel out when averaged over the whole neutral set, so that $\phi_{qp} \sim f_q$, while in the former case the local correlations do not cancel out at all because robustness is fundamentally a local quantity.

 It is quite striking that in all three models, a very large number of phenotypes are indeed connected to one another.  The HP model merits further discussion in this regard. In a recent review~\cite{goldstein2008structure},  RNA space was compared  to ``a bowl of spaghetti'', because the neutral spaces were connected to most other phenotypes,  while proteins were compared  to a``plum pudding'', where the neutral networks were more likely to be isolated from one another.
  We indeed find that the neutral networks in the HP24 model are not well connected,  but locate the origin of this effect in the large $N_P$/$N_G$ ratio for HP24, which means that many networks are below the threshold of Eq.~(\ref{eq:sample_threshold}) for connections.  By contrast, the compact HP5x5 model with many fewer phenotypes but a similar sized genotype space is well connected, more like ``spaghetti'' than like a ``plum pudding''.  What happens for real proteins, without the simplifying assumptions and small system sizes typically studied in the HP model~\cite{schram2013exact}, remains an open question.

Another type of heterogeneity in the mapping of genotypes to phenotypes can be quantified as {\em local} non-neutral correlations, which occur when the local neighbourhood of genotypes are different from the global expectation given by $\phi_{qp}$ or $f_q$.   We investigated two types of correlation (although one could imagine many more): i) non-neutral local over-representation correlations which result in phenotypes being more likely to be found multiple times around genotypes, and ii) non-neutral local mutational neighbourhood correlations, which mean that two genotypes connected by a neutral point mutation have mutational neighbourhoods that are more similar than do two randomly selected genotypes in a neutral set. 

These two types of correlation mean that the diversity of phenotypes in the direct neighbourhood of a genotype is lower than expected from the random model or even  from the averaged phenotype mutation coefficients $\phi_{qp}$.  Thus the rate at which a neutrally exploring population encounters novel variation will be reduced due to these correlations.  How this effect influences evolvability is complex, because the term is used in many different ways in the literature~\cite{conrad1990geometry,wagner1996perspective, Kirschner_1998,dawkins1989evolution,wagner2005robustness,pigliucci2008evolvability}. One type of evolvability simply measures the number of different phenotypes that are connected by single mutations to a neutral set~\cite{wagner2008robustness}. While  non-neutral correlations may not affect this number very much, they will affect the rate at which neutral exploration finds these new phenotypes. This lowering of the rate at which novelty appears may have a larger impact on other measures of evolvability. 

Each of the three models has a deleterious phenotype which either does not fold (for RNA and the HP protein model) or does not properly assemble (in the Polyomino model for protein clusters). The third type of non-neutral correlations we considered were iii) non-neutral deleterious phenotype correlations. For all three GP maps, the folding or assembling phenotypes have fewer mutational connections to the deleterious phenotypes than would be expected by the global frequency  $f_{\textrm{del}}$. This last result is perhaps the most interesting type of non-neutral correlation.  It was already predicted by John Maynard Smith in his classic 1970 paper~\cite{smith1970natural}, where he argued that ''meaningful" proteins were more likely to be neighbours of other "meaningful" proteins, and by extension,  that the probability of finding a deleterious phenotype in the mutational neighbourhood of a "meaningful" protein would be less than by random chance. 
 Such an effect can enhance evolutionary dynamics, because non-deleterious phenotypes are more strongly connected by mutations than expected by random chance, and so the population can more easily access potentially meaningful novel variation. Of course in practice, whether or not even the folding or self-assembling phenotypes are in fact ``meaningful'' will depend on the environment and other factors, but to first order a reduced propensity to mutate to manifestly deleterious phenotypes should be an evolutionary advantage.

\textbf{Evolvability:} While the effect of neutral correlations on robustness is straightforward, how correlations affect evolvability is more complex, not just because the concept itself is more diffuse, but also because the  relationships between correlations and evolvability  are more varied.  Nevertheless, we can summarise how different correlations affect evolvability as follows:

\begin{enumerate}
\item{Positive neutral correlations, measured by the presence of greater phenotypic robustness than would be expected by chance, is critical for the formation of large neutral networks. These networks are, in turn, a key facilitating factor for the ability of a population to access novel variation by neutral evolution over the network~\cite{smith1970natural,lipman1991modelling,schuster1994sequences,fontana2002modelling,cowperthwaite2008ascent,aguirre2011topological,schaper2014arrival,wagner2005robustness,wagner2011origins,schultes2000one,hayden2011cryptic,payne2014robustness}. Without positive neutral correlations, and the associated phenotype robustness, evolvability would be hugely suppressed.}
\item{The non-folding or non-assembling deleterious set of phenotypes are (positively) neutrally correlated and anti-correlated with the set of folding or assembling phenotypes.  This correlation increases the potential phenotypic variation that is accessible by reducing the likelihood of mutations leading to a seriously deleterious phenotype.}
\item{Local non-neutral correlations generally mean that the amount of novel variation available after mutations is smaller than one might expect from a random model.  These correlations will reduce evolvability.  For example,  Huynen et al.~\cite{huynen1996neutrality}  showed that the innovation rate for a random walk on the neutral space of the $L = 76$ secondary structure of tRNA$^{Phe}$ is about $20$, even though each genotype has $3L = 228$ neighbours. This significant lowering of the innovation rate is due to local non-neutral correlations, and may be a more general effect.
}
\item{For all  three GP maps we found that  $\phi_{qp} \approx f_p$, which means that the exponential variation in the frequency of phenotypes is reflected in the probability that a novel phenotype is found by mutations.  As argued in  ref.~\cite{schaper2014arrival},  for a range of evolutionary scenarios,  phenotypes with large $f_p$ will be exponentially more easy to find  than phenotypes with smaller $f_p$.   These large differences in the rate at which novel variation arrives will strongly modulate the evolvability.}
\end{enumerate}

\textbf{Other measures of genetic correlations:}
While we have built upon and introduced a number of metrics for genetic correlations in GP maps, the framework of correlations has other avenues for future computational work. For example, the central focus here has been on the way phenotypes are neutrally connected and non-neutrally connected without any broader concern for the properties of the phenotypes themselves. Phenotypes themselves have measurable properties such as symmetry, size and modularity and one could take the analysis further by considering whether there is a relationship between the mutational connections between two phenotypes and their similarity based on such properties. Making use of the null model again, where there is no predisposition for which phenotypes are mutationally close together, such \textit{phenotype similarity correlations} could be studied in the biological GP maps.

\textbf{Measuring correlations in experiment:}
Computational studies on theoretical systems ultimately need to be backed up with empirical evidence in real biological systems. Robustness to mutations in protein tertiary structure has been a well-studied area in this regard, with both mutagenesis and phylogenetic experiments being  used to illustrate robustness \cite{wagner2005robustness}. It may be that non-neutral local correlations could  be verified using mutagenesis experiments.  For example, for mutational neighbourhood correlations, two neighbouring genotypes both with a chosen structure  could have their neighbourhoods examined for the range of phenotypes and compared to the neighbourhood of a more distant genotype with the same methodology used here computationally, potentially replicating the findings Fig. \ref{fig7}, but for real molecules.  It may also be possible to measure the neutral correlation length $n^*$ by doing multiple mutation experiments.. However, because it is hard in practice to extract full GP map properties such as $f_p$ for a given phenotype, the most challenging aspect of such an experiment would be in generating an appropriate effectively random expectation. This same challenge holds for the other kinds of correlations we investigate in this paper.

A few final caveats are in order. In these models it is natural to use a restricted definition of a neutral mutation leading to exactly the same phenotype, whereas a more complete theory would count all mutations that are not visible to selection as effectively neutral.  Thus the full picture of how these correlation affect evolutionary dynamics is complex, and depends not just on the GP map itself, but more generally on the genotype to phenotype to fitness map, for which the environment plays a key role. Moreover, population genetic parameters such as the population size and the mutation rate must be taken into account. But notwithstanding these complications,  the important influence that structure in the GP map, in this case measured through the lens of genetic correlations, has on the manner in which variation arises (the ``arrival of the fittest"~\cite{wagner2014arrival}), and so on evolutionary dynamics,  should be evident,  confirming Maynard Smith's suggestions from many years ago. It may even be that without these correlations, Darwinian evolution, and therefore life itself, would not have been possible.

\section{Acknowledgements}
The authors would like to thank Edward Rolls for his initial analysis of John  Maynard Smith's word game.
This work was funded under EP/P504287/1 by the Engineering and Physical Sciences Research Council (https://www.epsrc.ac.uk). SEA is supported by The Royal Society (https://royalsociety.org/).

\appendix

\section{Associated analytical results of the sampling threshold}
\label{S1_Text}
We begin with an analysis of the sampling threshold $\gamma$ from average phenotypes in a given GP map. We showed in Eq.~\eqref{eq:sample_threshold} that for effective sampling, we require
$$
f_q > \frac{1}{F_p (K-1) L}
$$
which may be expressed alternatively as
$$
f_p f_q > \frac{1}{K^L(K-1)L}
$$
The average phenotype frequency may be written down as
$$
\left< f \right>_\text{phenotype} = \frac{1}{N_P} \sum_i f_i = \frac{1}{N_P}
$$
or with respect to the genotype sampling distribution as
$$
\left< f \right>_\text{genotype} = \sum_i f^2_i 
$$
Substituting in the smaller of the two, the average phenotype frequency, and then considering the required threshold for effectively sampling phenotype $q$ from the average phenotype $p$, we find
$$
f_q > \frac{N_P}{K^L (K-1) L}
$$
For RNA, where empirical scaling values are known ($N_P\approx 1.5 \times L^{-\frac{3}{2}} 1.8^L$ \cite{Schuster_1994}), we can further write
$$
f_q \gtrsim 0.45^L\frac{1}{L^{\frac{5}{2}}}
$$
for a phenotype $q$ to be effectively sampled.

We can change the question of effective sampling to ask the conditions on $N_P$ for the average phenotype $q$ to be accessed from the average phenotype $p$. In these circumstances, we can see that
$$
N_P < \sqrt{K^L (K-1) L}
$$
for which we can see RNA satisfies to an increasing extent for increasing $L$, as $2^L3L -1.93^L$  monotonically increases with increasing $L$.

Finally, we can also write down the approximate fraction less than the average phenotype frequency that is accessible from an average phenotype, through expressing $f_q = \chi \left<f\right>_\text{phenotype}$, which leads to a threshold fraction
\begin{equation}
\chi = \frac{N_P^2}{K^L (K-1) L}
\end{equation}
Again, using RNA as an example system this leads to
$$
\chi \propto 0.81^L \frac{1}{L^4}
$$
for a given length of RNA, showing that an increasing fraction of phenotypes with frequencies below the average may be effectively sampled from the average phenotype.

\begin{figure*}[!t]	\label{S1_fig}
	\centering
	\includegraphics{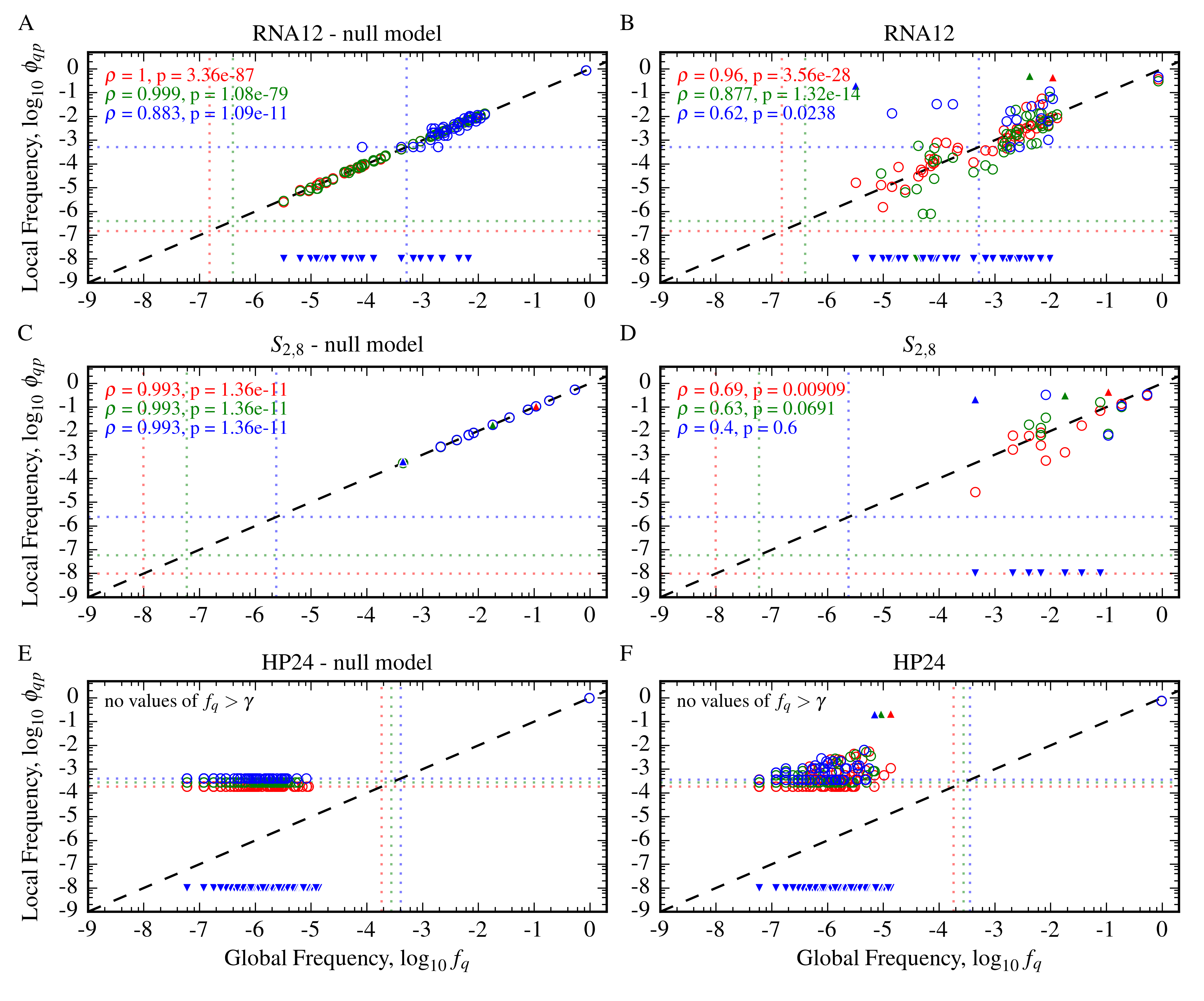}
	\caption{\textbf{The relationship between $\boldsymbol{\phi_{qp}}$ and $f_q$ is more complex in smaller GP maps.} Local frequency of phenotypes $q$ around genotypes with phenotype $p$ ($\phi_{qp}$) are plotted against the frequency of phenotypes $q$ ($f_q$) for each biological GP map and a random null model counterpart. The black dashed line is $\phi_{qp} = f_q$. The dotted lines are $\phi_{qp} = \gamma$ and $f_q=\gamma$ (c.f.  Eq.~\eqref{eq:sample_threshold}). In each case, three different phenotypes $p$ with different frequencies in the GP map are considered (represented with red, green and blue from largest to smallest $f_p$). (A, C and E) We plot local against global frequency for the random null models. $S_{2,8}$ and HP24 illustrate the two regimes where $f_q > \gamma$ and $f_q < \gamma$ in all cases respectively. The former has local frequencies strongly determined by global frequency ($\phi_{qp}=f_q$), while in the latter, occurrences of phenotypes are rare; they may not occur at all (downward triangular points, $\phi_{qp} = 0$) or they simply occur a single time ($\phi_{qp}=\gamma$). In the RNA12 null model, we see the blue phenotype crossing the threshold with some phenotypes having $f_q\approx \gamma$. (B, D, F) The three phenotypes are considered in each biological GP map. For larger frequency phenotypes (red and green in RNA12 and $S_{2,8}$), we find that local frequency is, to first order, well determined by the global frequency in line with the random null models (up to an order of magnitude variation in local frequency in comparison to global frequency). For lower frequency phenotypes (blue in RNA12 and $S_{2,8}$), we see that phenotype correlations are more important, an intuitive result given the genotypes of $p$ will be less encompassing of the whole GP map in these cases. In HP24 all frequencies are well below the gamma threshold but we still see a positive (although weaker) relationship between local frequency and global frequency (unlike in the null model, where $\phi_{qp}$ remains flat with respect to $f_q$ for $f_q<\gamma$). This is due to the presence of neutral correlations, an effect discussed in greater detail in the main text.}
\end{figure*}

\section{Extended analysis of $\boldsymbol{\phi_{qp}}$ across a broader range of phenotypes}
\label{S2_Text}
In this section, in contrast to the analysis in the main body considering the frequency of phenotypes $q$ around phenotype $p$ where $f_p\ll \gamma$, we consider different $p$ across a range of phenotype frequencies $f_p$ in the GP map.
In the random null model, at values of $f_q > \gamma$, we expect phenotypes to almost exactly follow $\phi_{qp} = f_q$. When $f_q <\gamma$, there are two likely possibilities for a given phenotype: either the phenotype is not found at all ($\phi_{qp}=0$) or it is found a single time ($\phi_{qp}=\gamma$). The latter is an over-representation of the local prevalence of $q$ for the GP map, while the former is clearly no local representation at all.

In Fig. 10, we display three pairs of plots for the RNA12, $S_{2,8}$ and HP24 GP maps and a randomised null model counterpart. The null models are displayed on the left, with the actual GP maps on the right. In each plot, we show the values of $\phi_{qp}$ against $f_q$ for three different phenotypes $p$ (coloured by data point as red, blue and green, with red the largest frequency phenotype and blue the smallest). Upward triangular data points represent values for $\phi_{pp}$, downward triangular data points $\phi_{qp}=0$ (shown at 1e-8 for visualisation purposes only) and the circular data points are all other phenotypes. Vertical and horizontal dotted coloured lines represent $f_q=\gamma$ and $\phi_{qp}=\gamma$ respectively. The diagonal dashed black line is $\phi_{qp}=f_q$, the null expectation for phenotypes with $f_q>\gamma$.

We begin by discussing the behaviour of the null model. The $S_{2,8}$ and HP24 null models provide the extreme cases. For $S_{2,8}$, all phenotypes are highly frequent and have $f_q \gg \gamma$. Consequently, we see that each of the three phenotypes follows the expected trend of $\phi_{qp}=f_q$ to a very high degree of accuracy (Spearman rank correlation coefficient and p-value in the top left). For the HP24 null model, all frequencies are such that $f_q \ll \gamma$. As such, phenotypes that are found locally are found only once ($\phi_{qp} = \gamma$) and most are not found at all (the many downward triangular points). For the RNA12 null model, the frequency of phenotypes used for phenotype $p$ span the range of all $f_q \gg \gamma$ (red and green) and to some phenotypes having $f_q\approx \gamma$ (blue). As a result, we see the red and green phenotypes follow $\phi_{qp} = f_q$ strongly, while the tail of the blue phenotype has fluctuations between the three behaviours.

The results from the null models demonstrate the accuracy of the above outlined intuition for a null relationship between the local connectivities of phenotypes with respect to the global abundance. With this in mind, we can now draw direct comparisons between each phenotype in the null model and actual behaviour exhibited in the biological GP maps. For each of the GP maps, we plot the same phenotype as in the null model case.
For RNA12, positive correlations are found for each phenotype, with deviations from $\phi_{qp}=f_q$ being more pronounced for lower frequency phenotypes (blue is subject to much greater fluctuations than red). The fluctuations are approximately up to an order of magnitude either side of the $\phi_{qp}=f_q$. We see a similar behaviour for $S_{2,8}$, with the largest fluctuations exhibited for the low frequency blue phenotype.

Finally, we consider the biological HP24 GP map. As was the case in the null model version, every phenotype (apart from the deleterious phenotype) lies in the region where $f_q \ll \gamma$. The notable difference in the actual GP map is the tendency for phenotypes with a larger $f_q$ to also be more likely to be locally present ($\log \phi_{qp} \propto \log f_q$). We can understand this with the following rationale: due to the neutral correlations present, if a single genotype with phenotype $q$ is found locally, then it is also likely that other genotypes with phenotype $q$ will be local to genotypes with phenotype $p$. And due to this effect being more pronounced for phenotypes with a greater frequency ($\rho_p \propto \log f_p$, c.f. Fig.~\ref{fig2}), we also see this effect locally with increased $f_q$ resulting in a greater $\phi_{qp}$, leading to the positive proportionality between $\log \phi_{qp}$ and $\log f_q$ in the actual GP maps when compared to the null models.

\end{document}